\documentclass[aps,pra,superscriptaddress,twocolumn]{revtex4}

\usepackage{amsfonts}
\usepackage{amsmath}
\usepackage{graphicx} 
\usepackage{color}
\usepackage{amssymb}
\usepackage[normalem]{ulem}
\usepackage{siunitx}
\usepackage{upgreek}
\usepackage{braket}
\usepackage{bm}
\DeclareMathAlphabet{\mathbbold}{U}{bbold}{m}{n}

\begin{document}

\title{Spectroscopic footprints of quantum friction in nonreciprocal and chiral media}

\author{O. J. Franca}
\email{uk081688@uni-kassel.de}
\affiliation{Institut f\"ur Physik, Universit\"at Kassel, Heinrich-Plett-Stra\ss e 40, 34132 Kassel, Germany}

\author{Fabian Spallek}
\affiliation{Institut f\"ur Physik, Universit\"at Kassel, Heinrich-Plett-Stra\ss e 40, 34132 Kassel, Germany}

\author{Steffen M. Giesen}
\affiliation{Fachbereich Chemie, Philipps-Universit\"at Marburg, Hans-Meerwein-Str 4, Marburg 35032, Germany}

\author{Robert Berger} 
\affiliation{Fachbereich Chemie, Philipps-Universit\"at Marburg, Hans-Meerwein-Str 4, Marburg 35032, Germany}

\author{Kilian Singer} 
\affiliation{Institut f\"ur Physik, Universit\"at Kassel, Heinrich-Plett-Stra\ss e 40, 34132 Kassel, Germany}

\author{Stefan Aull}
\affiliation{Institut f\"ur Physik, Universit\"at Kassel, Heinrich-Plett-Stra\ss e 40, 34132 Kassel, Germany}

\author{Stefan Yoshi Buhmann}
\email{stefan.buhmann@uni-kassel.de}
\affiliation{Institut f\"ur Physik, Universit\"at Kassel, Heinrich-Plett-Stra\ss e 40, 34132 Kassel, Germany}

\begin{abstract}
We investigate how the quantum friction experienced by a polarizable atom moving with constant velocity parallel to a planar interface is modified when the latter consists of chiral or nonreciprocal media, with special focus on topological insulators. Macroscopic quantum electrodynamics is used to obtain the velocity-dependent Casimir--Polder frequency shift and decay rate. These results are a generalization to matter with time-reversal symmetry breaking. Our findings are illustrated by examining the nonretarded and retarded limits for five examples: a perfectly conducting mirror, a perfectly reflecting nonreciprocal mirror, a three-dimensional topological insulator, a perfectly reflecting chiral mirror and an isotropic chiral medium. Different asymptotic power laws are found for all these materials. Interestingly, two bridges between chirality and nonreciprocity through the frequency shift, which arise as a consequence of the magnetoelectric coupling. Specifically, the position-dependent Casimir--Polder frequency shift for the nonreciprocal case depends on a geometric magnetic field associated with photoionization of chiral molecules, while the Casimir--Polder depending on the velocity for the chiral case has the optical rotatory strength as the atomic response, and those for the nonreciprocal case depend on an analog of the optical rotatory strength. 
\end{abstract}

\maketitle

\section{Introduction}
Quantum friction is closely linked to macroscopic effects of quantum vacuum fluctuations, such as the Casimir force between neutral bodies \cite{Casimir,Milonni,Dalvit-Milonni-Roberts-Rosa,Woods-et-al} and the Casimir–Polder force between an atom and a surface \cite{Casimir-Polder}. When two uncharged polarizable bodies move at constant velocity relative to each other, they experience a dissipative force opposing their motion due to the exchange of Doppler-shifted virtual photons at zero temperature—an effect known as quantum friction \cite{Pendry 1, Volokitin-Persson, Pendry 2,Silverinha, Belen}. For flat, perfectly smooth surfaces, quantum friction resembles a Van der Waals attraction, arising from quantum-fluctuation-induced electric dipoles on one surface and their lagging image dipoles on the other \cite{Pendry 1}. It has also been predicted in particle-surface (Casimir–Polder) interactions \cite{Scheel-Buhmann, Intravaia-Behunin-DAR, Intravaia et al PRL, Zhao et al}, with studies analyzing velocity- and distance-dependent drag forces in three-dimensional bulk materials and, more recently, in graphene \cite{Graphene}.

Quantum friction has a quantum origin and the physics behind it is related to the quantum Cherenkov effect through the anomalous Doppler effect \cite{Frolov,Nezlin,Ginzburg 2, Maghrebi}, where real photons are extracted from the vacuum at the cost of the object's kinetic energy; they are absorbed and emitted by an atom, thus producing a fluctuating momentum recoil \cite{Intravaia-Buhmann}. Considering only an atomic translational motion, spin-zero photons are absorbed and reemitted which translates in an overall quantum frictional force that opposes the translational motion. In recent years, this process has been investigated in several scenarios \cite{Dedkov,Barton, Advances QF,Bennett-Buhmann} and its outstanding connection to nonequilibrium physics was recently figured out \cite{Intravaia et al 1}. Generalizing for atomic rotational degrees of freedom results in what is called quantum rolling friction \cite{Intravaia et al 2}. Note that the related phenomenon of Coulomb drag \cite{Coulomb-drag}, in which a current in one plate induces a voltage bias in another via the fluctuating Coulomb field, has been successfully demonstrated \cite{Coulomb-drag-exp}. 

In order to gain further insights into quantum friction, a different path has been taken in the works \cite{Silverinha, Bennett-Buhmann}. Particularly, in the article \cite{Bennett-Buhmann} by using the framework of macroscopic QED \cite{MQED, MQED2, Acta, QF book} the authors are able to provide testable quantities with direct experimental contrast in the context of the Casimir--Polder force leading to the prediction of a velocity-dependent quantum-vacuum effect. Specifically, this work reports atomic level shifts and rates of spontaneous decay for a zero-temperature neutral atom with dipole moment and nonrelativistic velocity moving next to a perfectly smooth macroscopic surface, which in principle can be measured. Also important efforts have recently been done to obtain a satisfactory description of quantum friction in all timescales regarding carefully all the typical approximations for this phenomenon for example Markovianity and linear response theory \cite{Klatt-Kropf-Buhmann}. 

Besides the fundamental research of quantum friction, it has also been studied in innovative setups such as the flow of water in carbon nanochannels, where quantum friction is the dominant friction mechanism for water on carbon-based materials \cite{Kakovine}. This phenomenon has been investigated in new kinds of materials like bidimensional topological materials \cite{Belen}. For instance, exposing quantum friction to a two-dimensional topological material enables an increase of two orders of magnitude in the quantum drag force with respect to conventional neutral graphene systems \cite{Belen}. So, the jump to three-dimensional topological insulators (TIs) will open the gate to the interplay between quantum friction generated by an atom with constant velocity and the topological magnetoelectric effect provided by the TI, to which this paper is devoted.

Three-dimensional TIs are materials with an insulating bulk and conducting surface states protected by time-reversal symmetry \cite{Hasan,Qi Review,TIs Prediction 1,TIs Prediction 2}. Breaking time-reversal symmetry at a TI-insulator interface—via magnetic coatings or doping—opens a gap in surface states, leading to exotic phenomena \cite{Hasan,QAH,Mogi,Wu,Dziom,Qi PRB, Okada,OJF-LFU-ORT, OJF-SYB, OJF-LFU,OJF-SYB2,Qi Science}. TIs also show promise in hybrid systems, such as semiconductor quantum dots \cite{Castanho and us}, highlighting their vast potential for future applications. These effects can also occur in axion insulators, whose topological properties are protected by inversion symmetry and arise in magnetically doped TI heterostructures with opposing magnetization at top and bottom interfaces \cite{AXIs}.

Due to the fact that TIs  break time-reversal symmetry, they can be regarded as nonreciprocal media \cite{Buhmann-Butcher-Scheel, Fuchs-Crosse-Buhmann}. Consequently, the theory of macroscopic QED cannot be directly applied because it is based on Lorentz's reciprocity principle (a particular case of the Onsager reciprocity from statistical physics \cite{Onsager}), which assumes the reversibility of optical paths. In terms of symmetry, it is equivalent to invariance under an exchange of positions and orientations of sources and fields. Thus, media respecting this principle are called reciprocal and preserve time-reversal symmetry. To deal with nonreciprocal media, which violate Lorentz's principle relation \cite{Lorentz}, macroscopic QED was generalized to include the necessary features of nonreciprocal media in Ref.~\cite{Buhmann-Butcher-Scheel}. The aim of the present paper is to study the Casimir--Polder frequency shift and decay rate on a parallel moving atom to a nonreciprocal medium.

Another scenario where quantum friction has been investigated is in presence of chiral media as Ref.~\cite{Buhmann-Singer}, to which this paper is devoted as well. A three-dimensional object that cannot be superimposed on its mirror image is said to be chiral, these distinct mirror images are called enantiomers. Spectroscopically, enantiomers have identical transition energies and distinguishing between the two is not trivial. The characteristic feature of chiral objects is the manner of their interactions with other chiral objects. For example, the refractive indices of left- and right-handed circularly polarized light are different in a chiral medium, therefore the two polarizations will propagate at different speeds. The difference in velocity is related to the phenomenon of optical activity. Circular dichroism is closely related where the wave with the `slower' polarization is absorbed more strongly as it travels through the medium \cite{Condon}.

Many of the processes crucial to life involve chiral molecules whose chiral identity plays a central role in their chemical reactions. The undesired enantiomer can react differently, thereby not producing the required result. In nature, these chiral molecules predominantly, or even exclusively, occur in one enantiomeric form, so that their reactions proceed differently when different enantiomers are present. In contrast, artificial production without addition of enriched enantiomers or other chiral agents creates both enantiomers in equal proportions. It is essential to distinguish between enantiomers, and ultimately, to separate a racemic mixture --one that contains both enantiomers-- into an enantiomerically pure sample.

A frequently used method to separate enantiomers in an industrial setting is enantioselective chromotography. The initial racemic solution is passed through a column packed with a resolving agent, which by necessity has to be chiral. This either retards, or stops, the progress of one of the enantiomers passing through the column but crucially not for the other and thus allows the solution to be separated. From an optical perspective, it has been proposed that a racemic sample can be purified using coordinated laser pulses. This method involves a two-step process, where different transitions in the enantiomers are first driven, followed by a conversion of one enantiomer into the other. \cite{Kral}. It has recently been calculated that in the presence of a chiral carbon nanotube the enantiomers of alanine possess different absorption energies and it was theorized that this could lead to a method of discrimination \cite{Vardanega}. Furthermore, it has been shown that the van der Waals dispersion force between molecules can be enantiomer selective \cite{Hornberger, Molecular QED}. In fact, there have been a lot of efforts to provide a selection and distinguishing method such as the use of the Casimir--Polder force \cite{Butcher-Buhmann-Scheel,Barcellona,Suzuki}. For all these reasons, we will also explore how the quantum friction is modified in these materials. 



This paper is organized as follows: in Section \ref{EM FIELD} we calculate the time-dependent electric and magnetic field in the framework of macroscopic QED for nonreciprocal media. This result is obtained through a direct quantization of the noise current. Section \ref{INTERNAL} is devoted to study how the internal atomic dynamics is modified by the nonreciprocal media. To this end,  we present modified equations for the frequency shift and decay rate that depend on the atom position as well as on its velocity. In Section \ref{APPS} we apply our results to a electrically neutral polarizable atom moving with constant velocity parallel to five different materials: a perfectly conducting mirror, a perfectly reflecting nonreciprocal mirror, a strong three-dimensional topological insulator, a perfectly reflecting chiral mirror and an isotropic chiral medium. In Section \ref{DISCUSSION} we provide numerical results for the frequency shift with electrical dipole transition via Rydberg states and discuss the viability of an experimental verification. Section \ref{CONCLUSIONS} comprises a concluding summary of our results.

\section{ The time-dependent electromagnetic field } \label{EM FIELD}
Due to time-reversal symmetry violation in nonreciprocal media, the Lorentz reciprocity principle for the Green's tensor  \cite{Lorentz} is no longer valid
\begin{equation}
\mathbb{G}(\mathbf{r},\mathbf{r}';\omega) \neq \mathbb{G}^{ \top }(\mathbf{r}',\mathbf{r};\omega) \;
\end{equation}
where $^\top$ denotes the matrix transpose. This requires to introduce generalised definitions for the real and imaginary parts of the Green's tensor $\mathbb{G}$
\begin{eqnarray}
\mathcal{R}\left[ \mathbb{G}(\mathbf{r},\mathbf{r}';\omega) \right] &=& \frac{1}{2} \left[ \mathbb{G}(\mathbf{r},\mathbf{r}';\omega) + \mathbb{G}^{ *\top }(\mathbf{r}',\mathbf{r};\omega) \right] \;, \\
\mathcal{I}\left[ \mathbb{G}(\mathbf{r},\mathbf{r}';\omega) \right] &=& \frac{1}{2 \mathrm{i} } \left[ \mathbb{G}(\mathbf{r},\mathbf{r}';\omega) - \mathbb{G}^{ *\top }(\mathbf{r}',\mathbf{r};\omega) \right] \;. \label{Im}
\end{eqnarray}
For a reciprocal medium with
\begin{equation}\label{Lorentz reciprocity}
\mathbb{G}(\mathbf{r},\mathbf{r}';\omega) = \mathbb{G}^{ \top }(\mathbf{r}',\mathbf{r};\omega)\;,
\end{equation}
they reduce to ordinary, component-wise real and imaginary parts. The general expressions for the field operators read
\begin{eqnarray}
\mathbf{ \hat{E} } (\mathbf{r}) &=& \int_0^\infty d\omega \left[ \mathbf{ \hat{E} } (\mathbf{r};\omega) + \mathbf{ \hat{E} }^\dagger (\mathbf{r};\omega) \right] \;, \label{E r general} \nonumber\\
\mathbf{ \hat{B} } (\mathbf{r}) &=& \int_0^\infty d\omega \left[ \mathbf{ \hat{B} } (\mathbf{r};\omega) + \mathbf{ \hat{B} }^\dagger (\mathbf{r};\omega) \right] \;. \label{B r general}
\end{eqnarray}
where $^\dagger$ denotes hermitian conjugation in operator space which is distinct from $^{\top\ast}$, i.e.\ complex conjugation and matrix transpose in three-dimensional position space. The frequency components of the electric field (\ref{E r general}) can be computed through the Green's tensor in the following way
\begin{eqnarray}
\mathbf{ \hat{E} } (\mathbf{r};\omega) &=& \mathrm{i} \mu_0 \omega \left[ \mathbb{G} \star \mathbf{ \hat{j} }_N \right] (\mathbf{r};\omega) \;, \nonumber\\
&=& \mathrm{i} \mu_0 \omega \int d^3 r' \mathbb{G}(\mathbf{r},\mathbf{r}';\omega) \cdot  \mathbf{ \hat{j} }_N(\mathbf{r}';\omega) \;, \label{E GF j N}
\end{eqnarray}
where $\star$ denotes spatial convolution and $\mathbf{ \hat{j} }_N$ is the noise current density governed by the quantum fluctuations occurring in the medium and with a vanishing average $\langle\, \mathbf{ \hat{j} }_N \rangle = \mathbf{ 0 }$. The corresponding expression for the frequency components of the magnetic field (\ref{B r general}) is obtained by using Faraday's law in the frequency space
\begin{equation}\label{B GF j N}
\mathbf{ \hat{B} } (\mathbf{r};\omega) = \mu_0  \int d^3 r' \nabla \times \mathbb{G}(\mathbf{r},\mathbf{r}';\omega) \cdot  \mathbf{ \hat{j} }_N(\mathbf{r}';\omega) \;.
\end{equation}

We will directly quantize the noise current density $\mathbf{ \hat{j} }_N$ following the noise-current-based schema, as we shall expose in Sec. \ref{NOISE}. The Hamiltonian $\hat{ H }$ for the atom--field system is composed of the atomic part $\hat{ H }_A$, the field part $\hat{ H }_F$, and a contribution for the atom--field interaction $\hat{ H }_{AF}$, meaning that $\hat{ H }= \hat{ H }_A+\hat{ H }_{AF}+ \hat{ H }_F$. The atomic part is given by 
\begin{equation} \label{Hamiltonian A}
\hat{ H }_A = \sum_n E_n \hat{ A }_{nn} \;, 
\end{equation}
which incorporates the eigenenergies $E_n$ for each atomic energy level and the atomic flip operators $\hat{ A }_{mn}=|m\rangle \langle n|$. The Hamiltonian $\hat{ H }_F$ comprises the integral over all the frequency-dependent number operators of the field-medium system
\begin{equation} \label{Hamiltonian F}
\hat{ H }_F = \int d^3 \mathbf{r} \int_0^\infty d\omega \hbar \omega \mathbf{ \hat{f} }^\dagger (\mathbf{r};\omega) \cdot \mathbf{ \hat{f} } (\mathbf{r};\omega) \;,
\end{equation}
which clearly resembles a quantum harmonic oscillator and can be cast through the quantization of the noise current; cf. Sec. \ref{NOISE}. The interaction Hamiltonian $\hat{ H }_{AF}$ for nonmagnetic atoms in multipolar coupling and long wavelength approximation couples the atomic dipole to the electromagnetic field as 
\begin{eqnarray}
\hat{ H }_{AF} &=& -  \mathbf{ \hat{d} } \cdot \mathbf{ \hat{E} }(\mathbf{r}_A) + \mathbf{ \dot{ r } }_A \cdot \mathbf{ \hat{d} } \times \mathbf{ \hat{B} }(\mathbf{r}_A) \;, \nonumber\\
&=& - \sum_{m,n} \hat{ A }_{mn} \left[ \mathbf{d}_{mn} \cdot \mathbf{ \hat{E} }(\mathbf{r}_A) 
+ \mathbf{v} \cdot \mathbf{d}_{mn} \times \mathbf{ \hat{B} }(\mathbf{r}_A) \right] \;, \label{Hamiltonian AF} \nonumber\\
\end{eqnarray}
and contains the electric-dipole operator $\mathbf{ \hat{d} }= \sum_{m,n} \mathbf{d}_{mn} \hat{ A }_{mn}$, $\mathbf{r}_A$ is the position of the atom and $\mathbf{ \dot{ r } }_A=\mathbf{v}$ its velocity. As the atomic Hamiltonian $\hat{ H }_A$ and the field operators commute, only the resulting commutation relations for these operators with the field Hamiltonian $\hat{ H }_F$ and the interaction Hamiltonian $\hat{ H }_{AF}$ deserve to be studied to find the expression for the electric field (\ref{E r general}).\\

\subsection{Electromagnetic field in terms of noise currents} \label{NOISE}
In this subsection, we solve the equations of motion for the time-dependent creation $\mathbf{ \hat{f} }^\dagger$ and annihilation $\mathbf{ \hat{f} }$ operators, enabling us to determine the time-dependent electromagnetic field. For this aim, the noise current is quantized directly by expressions for the field operators. In frequency space, Ohm's law reads 
\begin{equation}
\mathbf{ \hat{ j } }(\mathbf{r};\omega) = \left[ \mathbb{Q} \star \mathbf{ \hat{E} } \right] (\mathbf{r};\omega) + \mathbf{ \hat{j} }_N(\mathbf{r};\omega) \;,
\end{equation}
which describes the response of a linear medium with conductivity matrix $\mathbb{Q}$ to the electric field $\mathbf{ \hat{E} } (\mathbf{r};\omega)$. Therefore, we write down the Helmholtz equations as follows
\begin{eqnarray}
&&\left[ \nabla \times \nabla - \frac{ \omega^2 }{ c^2 }\right] \mathbb{G}(\mathbf{r},\mathbf{r}';\omega) - \mathrm{i} \mu_0 \omega \left[ \mathbb{Q} \star \mathbb{G} \right] (\mathbf{r},\mathbf{r}';\omega) \nonumber\\
&& = \delta(\mathbf{r} - \mathbf{r}')\mathbbold{1}\;.
\end{eqnarray}
One finds the formal solution for this equation by employing the boundary condition $\mathbb{G}(\mathbf{r},\mathbf{r}';\omega)\rightarrow\mathbbold{0}$ for $\| \mathbf{r}-\mathbf{r}' \|\rightarrow\infty$, which leads to Eq.~(\ref{E GF j N}).

Next, we quantize the noise-current density $\mathbf{ \hat{j} }_N$ introduced in Eq.~(\ref{E GF j N}) simply by writing it in terms of creation $\mathbf{ \hat{f} }^\dagger$ and annihilation $\mathbf{ \hat{f} }$ operators
\begin{equation} \label{jN R and f}
\mathbf{ \hat{j} }_N (\mathbf{r};\omega) = \sqrt{ \frac{ \hbar\omega }{ \pi } } \left[ \mathbb{R} \star \mathbf{ \hat{f} } \right] (\mathbf{r};\omega)\;,
\end{equation}
where $\mathbb{R}$ is a square root of the positive definite conductivity tensor and is related to it by \cite{Buhmann-Butcher-Scheel}
\begin{equation} \label{R R dagger Q}
\left[ \mathbb{R} \star \mathbb{R}^{ *\top } \right] (\mathbf{r},\mathbf{r}';\omega) = \mathcal{R}\left[ \mathbb{Q}(\mathbf{r},\mathbf{r}';\omega) \right] \;.
\end{equation}

With all the above, we obtain the Heisenberg equation of motion for the annihilation operator $\mathbf{ \hat{f} }$ as shown
\begin{equation}
\mathbf{ \dot{ \hat{f} } } (\mathbf{r};\omega) = \frac{ 1 }{ \mathrm{i} \hbar } \left[ \mathbf{ \hat{f} }(\mathbf{r};\omega) , \hat{H} \right]\;, 
\end{equation}
where the three summands of $\hat{H}$ are given by Eqs.~(\ref{Hamiltonian A}), (\ref{Hamiltonian F}) and (\ref{Hamiltonian AF}). Nevertheless, it proves convenient to rewrite the interaction Hamiltonian $\hat{ H }_{AF}$ in terms of field operators, 
\begin{eqnarray}
&& \hat{ H }_{AF} = - \mathrm{i} \mu_0  \sum_{m,n}\int_0^\infty \omega d\omega \sqrt{ \frac{ \hbar\omega }{ \pi } } \hat{ A }_{mn} \nonumber\\
&& \times \mathbf{d}_{mn} \cdot \left\{ \left[ \mathbb{G} \star \mathbb{R} \star \mathbf{ \hat{f} } \right] (\mathbf{r}_A;\omega) - \left[ \mathbb{G}^* \star \mathbb{R}^* \star \mathbf{ \hat{f} }^\dagger \right] (\mathbf{r}_A;\omega) \right\} \nonumber\\
&& + \mu_0  \sum_{m,n}\int_0^\infty \frac{ d\omega }{ \mathrm{i}\omega } \sqrt{ \frac{ \hbar\omega }{ \pi } } \hat{ A }_{mn} \nonumber\\
&& \times \mathbf{v} \cdot \mathbf{d}_{mn} \times 
\left\{ \nabla \times \left[ \mathbb{G} \star \mathbb{R} \star \mathbf{ \hat{f} } \right] (\mathbf{r}_A;\omega) \right. \nonumber\\
&& \left. - \nabla \times \left[ \mathbb{G}^* \star \mathbb{R}^* \star \mathbf{ \hat{f} }^\dagger \right] (\mathbf{r}_A;\omega) \right\} \;.
\end{eqnarray}
In this way, the solution for the annihilation operator $\mathbf{ \hat{f} }$ is given by 
\begin{eqnarray}
&&\mathbf{ \hat{f} }(\mathbf{r};\omega,t) = \mathrm{e}^{ -\mathrm{i}\omega( t-t_0 ) }\mathbf{ \hat{f} }(\mathbf{r};\omega) + \frac{ \mu_0 \omega }{ \hbar } \sqrt{ \frac{ \hbar\omega }{ \pi } } \sum_{m,n} \int_{t_0}^t dt'  \nonumber\\
&& \times \left\{ \mathrm{e}^{ -\mathrm{i}\omega( t-t' ) } \left[ \mathbb{G} \star \mathbb{R}\right]^{ *\top }(\mathbf{r}_A (t'),\mathbf{r};\omega) \cdot  \mathbf{d}_{mn}  \hat{ A }_{mn}(t') \right. \nonumber\\
&& +  \frac{\mathrm{e}^{ -\mathrm{i}\omega( t-t' ) } }{ \omega^2 } \left( \left[ \mathbb{G} \star \mathbb{R}\right]^{ *\top }(\mathbf{r},\mathbf{r}_A(t');\omega)\times \overleftarrow{\nabla} \right) \nonumber\\
&& \times \left. \mathbf{d}_{mn} \cdot \mathbf{v} \hat{ A }_{mn}(t') \right\} \;, \label{Gen Annihilation}
\end{eqnarray}
where one should notice that $\overleftarrow{\nabla}$ acts on the second argument of the Green's tensor. Here we have required that $\mathbf{ \hat{f} }(\mathbf{r};\omega,t_0)=\mathbf{ \hat{f} }(\mathbf{r};\omega)$, i.e. the Heisenberg-picture operator agrees with its time-dependent Schr\"odinger-picture at initial time $t_0$.

Assuming that the atom moves with nonrelativistic speed, $v\ll c$, will enforce us to seek a solution to the atom--field dynamics that is correct within leading linear order of $v/c$. According to the Born--Oppenheimer approximation, we may assume the center of mass velocity to remain constant on the time-scale relevant for the internal atom--field dynamics, so that \cite{QF book}
\begin{equation}
\mathbf{r}_A(t')=\mathbf{r}_A(t)+(t-t')\mathbf{v} \;.
\end{equation}
Once this approximation is substituted into the solution for the annihilation operator (\ref{Gen Annihilation}) and discarding quadratic terms in $v/c$, we find that 
\begin{eqnarray}
&&\mathbf{ \hat{f} }(\mathbf{r};\omega,t) = \mathrm{e}^{ -\mathrm{i}\omega( t-t_0 ) }\mathbf{ \hat{f} }(\mathbf{r};\omega) + \frac{ \mu_0 \omega }{ \hbar } \sqrt{ \frac{ \hbar\omega }{ \pi } } \sum_{m,n} \int_{t_0}^t dt'  \nonumber\\
&& \times \left\{ \mathrm{e}^{ -\mathrm{i}\omega( t-t' ) } \left[ \mathbb{G} \star \mathbb{R}\right]^{ *\top }(\mathbf{r}_A (t),\mathbf{r};\omega) \cdot  \mathbf{d}_{mn}  \hat{ A }_{mn}(t') \right. \nonumber\\
&& - (t-t')\mathrm{e}^{ -\mathrm{i}\omega( t-t' ) } \left[ \mathbb{G} \star \mathbb{R}\right]^{ *\top }(\mathbf{r}_A (t),\mathbf{r};\omega) \nonumber\\
&&\cdot  \mathbf{d}_{mn} ( \overleftarrow{\nabla} \cdot \mathbf{v} ) \hat{ A }_{mn}(t') \nonumber\\
&& +  \frac{\mathrm{e}^{ -\mathrm{i}\omega( t-t' ) } }{ \omega^2 } \left( \left[ \mathbb{G} \star \mathbb{R}\right]^{ *\top }(\mathbf{r},\mathbf{r}_A(t);\omega)\times \overleftarrow{\nabla} \right) \nonumber\\
&& \times \left. \mathbf{d}_{mn} \cdot \mathbf{v} \hat{ A }_{mn}(t') \right\} \;. \label{Approx Annihilation}
\end{eqnarray}
After substituting the Eq.~(\ref{Approx Annihilation}) into Eq.~(\ref{jN R and f}), employing Eq.~(\ref{R R dagger Q}) and simplifying through the next integral relation for the Green's tensor in nonreciprocal media \cite{Buhmann-Butcher-Scheel}
\begin{equation}
\mathcal{I} \left[ \mathbb{G} (\mathbf{r},\mathbf{r}';\omega) \right] = \mu_0 \omega \left[ \mathbb{G} \star \mathcal{R}\left[ \mathbb{Q} \right] \star \mathbb{G}^{ *\top } \right] (\mathbf{r},\mathbf{r}';\omega) \;,
\end{equation}
we finally obtain expressions for the electric 
\begin{eqnarray}
&&\mathbf{ \hat{E} } (\mathbf{r};\omega,t) = \mathrm{e}^{ -\mathrm{i}\omega( t-t_0 ) }\mathbf{ \hat{E} }(\mathbf{r};\omega) + \frac{ \mathrm{i}\mu_0 \omega^2 }{ \pi } \sum_{m,n} \int_{t_0}^t dt'  \nonumber\\
&& \times \left\{ \mathrm{e}^{ -\mathrm{i}\omega( t-t' ) } \mathcal{I} \left[ \mathbb{G}(\mathbf{r},\mathbf{r}_A(t);\omega) \right] \cdot  \mathbf{d}_{mn}  \hat{ A }_{mn}(t') \right. \nonumber\\
&&  -(t-t')\mathrm{e}^{ -\mathrm{i}\omega( t-t' ) } \mathcal{I} \left[ \mathbb{G}(\mathbf{r},\mathbf{r}_A(t);\omega) \right] \nonumber\\
&&\cdot  \mathbf{d}_{mn} ( \overleftarrow{\nabla} \cdot \mathbf{v} ) \hat{ A }_{mn}(t') \nonumber\\
&& + \frac{\mathrm{e}^{ -\mathrm{i}\omega( t-t' ) } }{ \mathrm{i}\omega } \left( \mathcal{I} \left[ \mathbb{G}(\mathbf{r},\mathbf{r}_A(t);\omega) \right] \times \overleftarrow{\nabla} \right) \nonumber\\
&& \times \left. \mathbf{d}_{mn} \cdot \mathbf{v} \hat{ A }_{mn}(t') \right\} \;, \label{E Field}
\end{eqnarray}
and magnetic fields
\begin{eqnarray}
&&\mathbf{ \hat{B} } (\mathbf{r};\omega,t) = \mathrm{e}^{ -\mathrm{i}\omega( t-t_0 ) }\mathbf{ \hat{B} }(\mathbf{r};\omega) + \frac{ \mu_0 \omega }{ \pi } \sum_{m,n} \int_{t_0}^t dt'  \nonumber\\
&& \times \left\{ \mathrm{e}^{ -\mathrm{i}\omega( t-t' ) } \nabla \times \mathcal{I} \left[ \mathbb{G}(\mathbf{r},\mathbf{r}_A(t);\omega) \right] \cdot  \mathbf{d}_{mn}  \hat{ A }_{mn}(t') \right\} \label{B Field} \nonumber\\
&& + \mathcal{O}(\mathbf{v}/c) \;, 
\end{eqnarray}
where we only considered the zero-order approximation in $\mathbf{v}/c$ because the magnetic field appears in conjunction with a factor $\mathbf{v}$. In the latter equation, we remark that $\nabla$ acts only on the first argument of the Green's tensor.

Brief observations on Eqs.~(\ref{E Field}) and (\ref{B Field}) should be given. First, Eq.~(\ref{E Field}) extends the result for an atom at rest in nonreciprocal media of the work \cite{Fuchs-Crosse-Buhmann} to a moving atom. Second, the Eq.~(\ref{B Field}) generalizes the standard magnetic field given by Eq.~(8.11) of Ref.~\cite{QF book} to the case of nonreciprocal media. Both expressions differ from the usual expressions for reciprocal media, for which Eq.~(\ref{Lorentz reciprocity}) applies, only by the redefinition of the imaginary part of the Green's tensor defined in Eq.~(\ref{Im}).

\section{ Internal atomic dynamics: Frequency shift and decay rate } \label{INTERNAL}
The internal atomic dynamics can be described by the Heisenberg equations of motion for the flip operator
\begin{equation}
\dot{ \hat{ A } }_{mn} = \frac{1}{ \mathrm{i} \hbar }\left[ \hat{ A }_{mn}, \hat{H} \right] = \frac{1}{ \mathrm{i} \hbar }\left[ \hat{ A }_{mn}, \hat{H}_A \right] + \frac{1}{ \mathrm{i} \hbar }\left[ \hat{ A }_{mn}, \hat{H}_{AF} \right] \;,
\end{equation}
which does not include the field Hamiltonian $\hat{H}_F$ because it commutes with the flip operator. Our approach follows the procedure for a reciprocal surface fully detailed in Ref.~\cite{QF book} and the one applied for a nonreciprocal surface with an atom at rest of Ref.~\cite{Fuchs-Crosse-Buhmann}, now extended to a moving atom parallel to the surface.

This leads to 
\begin{equation}
\begin{aligned}
& \dot{ \hat{ A } }_{mn} = \mathrm{i}\omega_{mn}\hat{ A }_{mn} + \frac{ \mathrm{i} }{ \hbar } \sum_{k}\int_0^\infty d\omega \nonumber\\
& \times \left\{ \left( \hat{ A }_{mk} \mathbf{d}_{nk} - \hat{ A }_{kn} \mathbf{d}_{km} \right) \cdot \left[\mathbf{ \hat{E} }(\mathbf{r}_A; \omega) + \mathbf{v} \times \mathbf{ \hat{B} }(\mathbf{r}_A; \omega) \right] \right. \nonumber\\
& + \left.  \left[\mathbf{ \hat{E} }^\dagger (\mathbf{r}_A; \omega) + \mathbf{v} \times \mathbf{ \hat{B} }^\dagger (\mathbf{r}_A; \omega) \right] \cdot \left( \mathbf{d}_{nk} \hat{ A }_{mk}  - \mathbf{d}_{km} \hat{ A }_{kn}  \right) \right\} , \label{Amn eom} \nonumber\\
\end{aligned}
\end{equation}
which appears in normal ordering and does not show the time-dependence for brevity. 

To evaluate the required time integrals, we will make some approximations. Firstly, we will assume weak atom--field coupling, which requires that the field spectrum $\omega^2 \mathcal{I} \left[ \mathbb{G}(\mathbf{r}_A(t),\mathbf{r}_A(t);\omega) \right]$ is sufficiently flat \cite{QF book}. Particularly, it must not exhibit any narrow peaks in the vicinity of any atomic transition frequency. Secondly, we will assume that $\hat{ A }_{mn}$ is dominated by oscillations with frequencies $\tilde{\omega}_{mn}$, which are yet to be determined in a self-consistent manner. Lastly, we will formally carry out the time integrals by means of the Markov approximation. Thus, we are allowed to neglect the slow non-oscillatory dynamics of the atomic flip operator $\hat{ A }_{mn}$ during the time interval $t'\in[t_0,t]$, so we may set $\hat{ A }_{mn}(t')\simeq\exp[ \mathrm{i} \tilde{\omega}_{mn} (t'-t) ]\hat{ A }_{mn}(t)$, where we foresee that the property $\tilde{\omega}_{mn}=-\tilde{\omega}_{nm}$ must be verified. Then, we send $t_0\rightarrow-\infty$ in the time integral obtaining 
\begin{eqnarray}
&& \hat{ A }_{mn}(t)\int_{t_0}^t dt' \mathrm{e}^{ -\mathrm{i}( \omega - \tilde{\omega}_{nm} )( t-t' ) } \nonumber\\
&& \simeq \hat{ A }_{mn}(t)\int_{-\infty}^t dt' \mathrm{e}^{ -\mathrm{i}( \omega - \tilde{\omega}_{nm} )( t-t' ) }\;, \nonumber\\
&& = \hat{ A }_{mn}(t) \left[ \pi \delta(\omega - \tilde{\omega}_{nm} ) - \mathrm{i} \frac{ \mathcal{P} }{ \omega - \tilde{\omega}_{nm} } \right]\;. \label{Markov 1}
\end{eqnarray}
By taking the derivative with respect to $\omega$ of this result, we determine the remaining time integral 
\begin{eqnarray}
&&\int_{t_0}^t dt' (t-t') \mathrm{e}^{ -\mathrm{i}\omega( t-t' ) } \hat{ A }_{mn}(t') \nonumber\\
&& \simeq \frac{d}{ d\omega } \left[ \frac{ \mathcal{P} }{ \omega - \tilde{\omega}_{nm} } + \mathrm{i} \pi \delta(\omega - \tilde{\omega}_{nm} ) \right] \hat{ A }_{mn}(t) \;. \label{Markov 2}
\end{eqnarray}
For both results $\mathcal{P}$ stands for the Cauchy principle value and the limits of the frequency integral of Eq.~(\ref{Amn Heisenberg}) will lead to the appearance of Heaviside function $\vartheta$.

By defining the coefficient 
\begin{equation}
\mathbf{C}_{mn} = \mathbf{C}_{mn} (\mathbf{r}_A) + \mathbf{C}_{mn} (\mathbf{r}_A,\mathbf{v}) \;,
\end{equation}
which contains a pure position-dependent component \cite{Fuchs-Crosse-Buhmann}
\begin{eqnarray}
 \mathbf{C}_{mn} (\mathbf{r}_A) &=& \frac{ \mu_0 }{ \hbar } \vartheta( \tilde{ \omega }_{mn} ) \tilde{ \omega }_{mn}^2 \mathcal{I}\left[ \mathbb{G}(\mathbf{r}_A,\mathbf{r}_A;\tilde{ \omega }_{mn}) \right] \cdot \mathbf{d}_{mn} \nonumber\\
&& - \frac{ \mathrm{i} \mu_0 }{ \pi \hbar } \mathcal{P} \int_0^\infty \frac{ \omega^2 d\omega }{ \omega - \tilde{\omega}_{nk} } \mathcal{I} \left[ \mathbb{G}(\mathbf{r}_A,\mathbf{r}_A;\omega ) \right] \cdot \mathbf{d}_{mn} , \label{Cmn r} \nonumber\\
\end{eqnarray}
as well as a position- and velocity-dependent one
\begin{equation}
\begin{aligned}
&\mathbf{C}_{mn} (\mathbf{r}_A,\mathbf{v}) =  \\
& \frac{ \mathrm{i} \mu_0 }{ \hbar } \vartheta( \tilde{ \omega }_{mn} ) \left\{ \omega^2 \mathcal{I} \left[ \mathbb{G}(\mathbf{r}_A,\mathbf{r}_A;\omega) \right] \right\}_{\omega=\tilde{\omega}_{nm}}^\prime \cdot \mathbf{d}_{mn} ( \overleftarrow{\nabla} \cdot \mathbf{v} ) \\
& + \frac{ \mu_0 }{ \pi \hbar } \mathcal{P} \int_0^\infty \frac{ d\omega }{ \omega - \tilde{\omega}_{nm} } \left\{ \omega^2 \mathcal{I} \left[ \mathbb{G}(\mathbf{r}_A,\mathbf{r}_A;\omega) \right] \right\}^\prime \cdot \mathbf{d}_{mn} ( \overleftarrow{\nabla} \cdot \mathbf{v} )  \\
& -  \frac{ \mathrm{i} \mu_0 }{ \hbar } \vartheta( \tilde{ \omega }_{mn} ) \tilde{ \omega }_{nm} \left\{ \mathcal{I} \left[ \mathbb{G}(\mathbf{r}_A,\mathbf{r}_A;\tilde{ \omega }_{nm}) \right] \times \overleftarrow{\nabla} \right\} \times \mathbf{d}_{mn} \cdot \mathbf{v} \\
& - \frac{ \mu_0 }{ \pi \hbar } \mathcal{P} \int_0^\infty \frac{ \omega d\omega }{ \omega - \tilde{\omega}_{nm} } \left\{ \mathcal{I} \left[ \mathbb{G}(\mathbf{r}_A,\mathbf{r}_A; \omega ) \right] \times \overleftarrow{\nabla} \right\} \times \mathbf{d}_{mn} \cdot \mathbf{v} \\
& -  \frac{ \mathrm{i} \mu_0 }{ \hbar } \vartheta( \tilde{ \omega }_{mn} ) \tilde{ \omega }_{nm} \mathbf{v} \times \left\{ \nabla \times \mathcal{I} \left[ \mathbb{G}(\mathbf{r}_A,\mathbf{r}_A; \tilde{ \omega }_{nm} ) \right]  \right\} \cdot \mathbf{d}_{mn} \\
& - \frac{ \mu_0 }{ \pi \hbar } \mathcal{P} \int_0^\infty \frac{ \omega d\omega }{ \omega - \tilde{\omega}_{nm} } \mathbf{v} \times \left\{ \nabla \times \mathcal{I} \left[ \mathbb{G}(\mathbf{r}_A,\mathbf{r}_A; \omega  ) \right]  \right\} \cdot \mathbf{d}_{mn} , \label{Cmn r v} \\
\end{aligned}
\end{equation}
where primes in the first and second term denote derivatives with respect to the respective frequency arguments. We are able to cast Eq.~(\ref{Amn eom}) into the form
\begin{eqnarray}
\dot{ \hat{ A } }_{mn}(t) &=& \mathrm{i} \omega_{mn} \hat{ A }_{mn}(t) + \frac{ \mathrm{i} }{ \hbar } \sum_{k}\int_0^\infty d\omega \nonumber\\ 
&& \times \left\{ e^{ -\mathrm{i} \omega ( t - t_0) } \left[ \hat{ A }_{mk}(t) \mathbf{d}_{nk} - \hat{ A }_{kn}(t) \mathbf{d}_{km} \right]  \right. \nonumber\\
&&  \cdot \left[\mathbf{ \hat{E} }(\mathbf{r}_A; \omega) + \mathbf{v} \times \mathbf{ \hat{B} }(\mathbf{r}_A; \omega) \right] \nonumber\\
&& + e^{ \mathrm{i} \omega ( t - t_0) } \left[\mathbf{ \hat{E} }^\dagger (\mathbf{r}_A; \omega) + \mathbf{v} \times \mathbf{ \hat{B} }^\dagger (\mathbf{r}_A; \omega) \right]  \nonumber\\
&&\left. \cdot \left( \mathbf{d}_{nk} \hat{ A }_{mk}(t)  - \mathbf{d}_{km} \hat{ A }_{kn}(t)  \right)\right\} \nonumber\\
&& -\sum_{k,l} \left[ \mathbf{d}_{nk} \cdot \mathbf{C}_{kl} \hat{ A }_{ml}(t) - \mathbf{d}_{km} \cdot \mathbf{C}_{nl} \hat{ A }_{kl}(t) \right] \nonumber\\
&& + \sum_{k,l} \left[ \mathbf{d}_{nk} \cdot \mathbf{C}_{ml}^* \hat{ A }_{lk}(t) - \mathbf{d}_{km} \cdot \mathbf{C}_{kl}^* \hat{ A }_{ln}(t) \right] \;, \label{Amn Heisenberg} \nonumber\\
\end{eqnarray}
where we have used the identity $\mathcal{I}[ \mathbb{G}^*(\mathbf{r}_A,\mathbf{r}_A;\omega) ] = \mathcal{I}[ \mathbb{G}^\top(\mathbf{r}_A,\mathbf{r}_A;\omega) ]$, which is deduced from Eq.~(\ref{Im}).  The main difference between Eq.~(\ref{Amn Heisenberg}) and the one for the reciprocal case, for which Eq.~(\ref{Lorentz reciprocity}) applies, lies in the redefinition of the imaginary part of the Green's tensor (\ref{Im}) introduced by the coefficients (\ref{Cmn r}) and (\ref{Cmn r v}), as can be contrasted with the coefficients for reciprocal media given by Eqs.~(5.52) and (8.17) reported on Ref.~\cite{QF book}. 

Next, we take expectation values of Eq.~(\ref{Amn Heisenberg}) assuming that the electromagnetic field is prepared in its ground state at initial time $t_0$, which implies that $\mathbf{ \hat{E} }(\mathbf{r}; \omega)|\{ 0 \} \rangle = \mathbf{0},$ $\forall \mathbf{r}, \omega$. This also means that the involved density matrix reads $\hat{ \rho }_0=|\{ 0 \} \rangle \langle \{ 0 \} | $. As a consequence of normal ordering the free contributions of the electric and magnetic field will be zero. 

Now, we need to solve Eq.~(\ref{Amn Heisenberg}). To this end, we assume the atom to be free of quasidegenerate transitions. Additionally, the atom is assumed to be unpolarized in each of its energy eigenstates, i.e. $\mathbf{d}_{nn}=\mathbf{0}$, as well as that its states of a degenerate manifold are not connected by electric-dipole transitions $\mathbf{d}_{nn'}=\mathbf{0}$. Both conditions are guaranteed by atomic selection rules \cite{QF book}. Under these conditions, the fast-oscillating off-diagonal flip operators decouple from the nonoscillating diagonal ones as well as from each other \cite{QF book}. 

For $m\neq n$, under the above assumptions, we obtain that 
\begin{eqnarray}
\langle \dot{ \hat{ A } }_{mn}(t) \rangle &=&  \mathrm{i} \omega_{mn}  \langle  \hat{ A }_{mn}(t) \rangle \nonumber\\
&& - \sum_k \left(\mathbf{d}_{nk} \cdot \mathbf{C}_{kn} + \mathbf{d}_{km} \cdot \mathbf{C}_{km}^*\right) \langle  \hat{ A }_{mn}(t) \rangle . \label{Amn ODE}  \nonumber\\
\end{eqnarray}

For $m=n$, the diagonal flip operators are non-oscillating and mutually coupled. After retaining only diagonal terms on Eq.~(\ref{Amn Heisenberg}), we find 
\begin{eqnarray}
\langle \dot{ \hat{ A } }_{nn}(t) \rangle &=& - \sum_k \left(\mathbf{d}_{nk} \cdot \mathbf{C}_{kn} + \mathbf{d}_{kn} \cdot \mathbf{C}_{kn}^*\right) \langle  \hat{ A }_{nn}(t) \rangle \nonumber\\
&& + \sum_k \left(\mathbf{d}_{kn} \cdot \mathbf{C}_{nk} + \mathbf{d}_{nk} \cdot \mathbf{C}_{nk}^*\right) \langle  \hat{ A }_{kk}(t) \rangle . \label{Ann ODE} \nonumber\\
\end{eqnarray}


By virtue of Eq.~(\ref{Im}) the two terms $\mathbf{d}_{nk} \cdot \mathcal{I}[ \mathbb{G}(\mathbf{r}_A,\mathbf{r}_A;\omega) ]\cdot \mathbf{d}_{kn} = \mathrm{Im}[ \mathbf{d}_{nk} \cdot \mathbb{G}(\mathbf{r}_A,\mathbf{r}_A;\omega) \cdot \mathbf{d}_{kn} ]$ and $\mathbf{d}_{kn}\cdot\mathcal{I}[ \mathbb{G}^{ \top }(\mathbf{r}_A,\mathbf{r}_A;\omega) ] \cdot \mathbf{d}_{nk} = \mathrm{Im}[ \mathbf{d}_{nk} \cdot \mathbb{G}(\mathbf{r}_A,\mathbf{r}_A;\omega)\cdot \mathbf{d}_{kn} ] $ are equal and real. This will occur for the real parts too. 

At this point it is worth mentioning that the Green's tensor $\mathbb{G}$ can be split into a bulk part $\mathbb{G}^{(0)}$ and a scattering part $\mathbb{G}^{(1)}$. Here we will assume that the Lamb shift associated with the free-space Green's tensor $\mathbb{G}^{(0)}$ is already included in the bare transition frequency $\omega_{mn}$ by making the replacement $\mathbb{G}\rightarrow\mathbb{G}^{(1)}$, which refers only to the atom and does not consider the material properties of surrounding matter \cite{MQED}. Hence, the remaining frequency shift stems from the presence of electromagnetic bodies around the atom.

So, taking real and imaginary parts of the coefficients in Eqs.~(\ref{Amn ODE}) and (\ref{Ann ODE}) according to
\begin{eqnarray}
\sum_k \mathbf{d}_{nk} \cdot \mathbf{C}_{kn} &=& \frac{ \Gamma_n }{2} + \mathrm{i}\delta\omega_n \;, \\
\sum_k \mathbf{d}_{kn} \cdot \mathbf{C}_{kn}^* &=& \frac{ \Gamma_n }{2} - \mathrm{i}\delta\omega_n \;,
\end{eqnarray}
we are able to identify the decay rate 
\begin{eqnarray}
\Gamma_n &=& \Gamma_n (\mathbf{r}_A) + \Gamma_n (\mathbf{r}_A, \mathbf{v}) \;, \nonumber\\
&=& \sum_{ k } \Gamma_{nk} = \sum_{ k<n } \Gamma_{nk} (\mathbf{r}_A) + \sum_{ k } \Gamma_{nk} (\mathbf{r}_A, \mathbf{v}) \, , \;\; \label{whole rate}
\end{eqnarray}
and the frequency shift 
\begin{eqnarray}
\delta\omega_n &=& \delta\omega_n (\mathbf{r}_A) + \delta\omega_n (\mathbf{r}_A, \mathbf{v}) \;, \nonumber\\
&=& \sum_k \delta\omega_{nk} = \sum_k \delta\omega_{nk} (\mathbf{r}_A) + \sum_{ k<n } \delta\omega_{nk} (\mathbf{r}_A, \mathbf{v}) , \;\;\;\;\;\;\; \label{whole shift}
\end{eqnarray}
%
with the position-dependent contributions \cite{Fuchs-Crosse-Buhmann}
\begin{equation}\label{Decay rate r}
\Gamma_{nk} (\mathbf{r}_A) =  \frac{ 2\mu_0 }{ \hbar } \tilde{\omega}_{nk}^2 \,\mathrm{Im} \left[ \mathbf{d}_{nk} \cdot \mathbb{G}^{(1)}(\mathbf{r}_A,\mathbf{r}_A;\tilde{\omega}_{nk}) \cdot \mathbf{d}_{kn} \right] \;,
\end{equation}
\begin{eqnarray}
\delta\omega_{nk} (\mathbf{r}_A) &=& - \frac{ \mu_0 }{ \pi \hbar } \mathcal{P} \int_0^\infty \frac{ \omega^2 d\omega }{ \omega - \tilde{\omega}_{nk} } \nonumber\\
&& \times \mathrm{Im} \left[ \mathbf{d}_{nk} \cdot \mathbb{G}^{(1)}(\mathbf{r}_A,\mathbf{r}_A;\omega ) \cdot \mathbf{d}_{kn} \right] \label{Freq shift r} \;, 
\end{eqnarray}
and the position- and velocity-dependent contributions
\begin{eqnarray}
&&\Gamma_{nk} (\mathbf{r}_A, \mathbf{v}) = \nonumber\\
&&\frac{ 2\mu_0 }{ \pi \hbar } \mathcal{P} \int_0^\infty \frac{ d\omega }{ \omega - \tilde{\omega}_{nk} } \nonumber\\
&& \times \left\{ \omega^2 \mathcal{I} \left[ \mathbf{d}_{nk} \cdot ( \mathbf{v} \cdot \nabla )\mathbb{G}^{(1)}(\mathbf{r}_A,\mathbf{r}_A;\omega) \cdot \mathbf{d}_{kn} \right] \right\}' \nonumber\\
&& - \frac{ 2\mu_0 }{ \pi \hbar } \mathcal{P} \int_0^\infty \frac{ \omega d\omega }{ \omega - \tilde{\omega}_{nk} } \nonumber\\
&& \times \mathbf{d}_{nk} \cdot \left\{ \mathcal{I} \left[ \mathbb{G}^{(1)}(\mathbf{r}_A,\mathbf{r}_A;\omega) \right] \times \overleftarrow{\nabla} \right\} \times \mathbf{d}_{kn} \cdot \mathbf{v} \nonumber\\
&& - \frac{ 2\mu_0 }{ \pi \hbar } \mathcal{P} \int_0^\infty \frac{ \omega d\omega }{ \omega - \tilde{\omega}_{nk} } \nonumber\\
&& \times \mathbf{d}_{nk} \cdot \mathbf{v} \times \left\{ \nabla \times \mathcal{I} \left[ \mathbb{G}^{(1)}(\mathbf{r}_A,\mathbf{r}_A;\omega) \right]  \right\} \cdot \mathbf{d}_{kn} \;, \label{Decay rate r v} 
\end{eqnarray}
\begin{eqnarray}
&&\delta\omega_{nk} (\mathbf{r}_A, \mathbf{v}) = \nonumber\\
&&\frac{ \mu_0 }{ 2\mathrm{i}\hbar } \left[ \omega^2  \mathbf{d}_{kn}\cdot ( \mathbf{v} \cdot \nabla' ) \mathbb{G}^{(1)\top}(\mathbf{r},\mathbf{r}';\omega) \cdot \mathbf{d}_{nk} \right]'_{\substack{\omega=\tilde{\omega}_{nk} \\ \mathbf{r}=\mathbf{r}'=\mathbf{r}_A }} \nonumber\\
&&-\frac{ \mu_0 }{ 2\mathrm{i}\hbar } \left[ \omega^2  \mathbf{d}_{nk}\cdot ( \mathbf{v} \cdot \nabla' ) \mathbb{G}^{(1)\top*}(\mathbf{r},\mathbf{r}';\omega) \cdot \mathbf{d}_{kn} \right]'_{\substack{\omega=\tilde{\omega}_{nk} \\ \mathbf{r}=\mathbf{r}'=\mathbf{r}_A }} \nonumber\\
&& - \frac{ \mu_0 }{ \hbar } \tilde{\omega}_{nk} \nonumber\\
&& \times \mathrm{Im} \left\{ \mathbf{v} \cdot \mathbf{d}_{kn} \times   \left[ \nabla \times \mathbb{G}^{(1)\top}(\mathbf{r}',\mathbf{r};\tilde{\omega}_{nk}) \right] \cdot \mathbf{d}_{nk} \right\}_{ \mathbf{r}=\mathbf{r}'=\mathbf{r}_A }    \nonumber\\
&& + \frac{ \mu_0 }{ \hbar } \tilde{\omega}_{nk} \nonumber\\
&& \times\mathrm{Im} \left\{ \mathbf{v} \cdot \mathbf{d}_{nk} \times   \left[ \nabla \times \mathbb{G}^{(1)}(\mathbf{r},\mathbf{r}';\tilde{\omega}_{nk}) \right] \cdot \mathbf{d}_{kn} \right\}_{ \mathbf{r}=\mathbf{r}'=\mathbf{r}_A } . \label{Freq shift r v} \nonumber\\
\end{eqnarray}

From Eqs.~(\ref{Decay rate r v}) and (\ref{Freq shift r v}) we observe that the second and third terms can be traced back to the last four terms of the coefficient $\mathbf{C}_{mn} (\mathbf{r}_A,\mathbf{v})$ defined in Eq.~(\ref{Cmn r v}). In fact, for an atom  with time-reversal invariant internal Hamiltonian and hence real dipole-matrix elements, such terms will cancel pairwise and will not contribute. However, in the present work we will take into account all these terms of $\mathbf{C}_{mn}(\mathbf{r}_A,\mathbf{v})$, because we will choose dipole moments such that the atom is sensitive to chiral media or the violated time-symmetry of the nonreciprocal media. Our choice is based on Curie's principle, which states that a system consisting of a crystal and an external influence, each having a specific symmetry, only maintains the symmetries that are shared by both the crystal and the external influence \cite{Curie}. 

With the definitions (\ref{Decay rate r v}) and (\ref{Freq shift r v}), the equations of motion (\ref{Amn ODE}) and (\ref{Ann ODE}) for the expectation values of the atomic flip operators take the final form
\begin{eqnarray}
\langle \dot{ \hat{ A } }_{mn}(t) \rangle &=& \left[ \mathrm{i} \tilde{\omega}_{mn} - \frac{1}{2}( \Gamma_m + \Gamma_n  )  \right] \langle  \hat{ A }_{mn}(t) \rangle \;, \label{Amn omega gamma ODE} \\
\langle \dot{ \hat{ A } }_{nn}(t) \rangle &=& -\Gamma_n \langle  \hat{ A }_{nn}(t) \rangle + \sum_{ k<n } \Gamma_{nk} \langle  \hat{ A }_{kk}(t) \rangle \;, \label{Ann omega gamma ODE}
\end{eqnarray}
where we have identified the frequencies 
\begin{equation}
\tilde{\omega}_{mn}=\omega_{mn}+\delta\omega_m-\delta\omega_n \;,
\end{equation}
which verifies the property $\tilde{\omega}_{mn}=-\tilde{\omega}_{nm}$ previously mentioned. The appearance of the shifted frequency $\tilde{\omega}_{nk}$ in the expression for $\delta\omega_{nk}$ through the contributions $\delta\omega_{nk} (\mathbf{r}_A)$ (\ref{Freq shift r}) and $\delta\omega_{nk} (\mathbf{r}_A, \mathbf{v})$ (\ref{Freq shift r v}) itself, the frequency shift is given as a self-consistent result from the implicit equation. 

Henceforth we restrict our analysis only on the frequency shifts (\ref{Freq shift r}) and (\ref{Freq shift r v}), which can be further simplified by using the redefinition of the imaginary part of the Green's tensor (\ref{Im}), the Schwarz principle valid for nonreciprocal media too, 
\begin{equation}
\mathbb{G}^*(\mathbf{r}_A,\mathbf{r}_A;\omega) = \mathbb{G}^*(\mathbf{r}_A,\mathbf{r}_A;-\omega^*) \;,
\end{equation}
and the substitution $\omega\rightarrow-\omega$ in their second integrals arising from Eq.~(\ref{Im}). The integral contours along the positive and negative axes, each has one real pole and is evaluated in the $\omega$-complex plane. The path along the quarter circle does not contribute because $\lim_{|\omega|\rightarrow0} \omega^2\mathbb{G}^{(1)}(\mathbf{r}_A,\mathbf{r}_A;\omega)/c^2=\mathbbold{0}$. The part along the the imaginary axis leads to the position-dependent nonresonant frequency shift \cite{Fuchs-Crosse-Buhmann}
\begin{eqnarray}
\delta\omega_{nk}^{ \mathrm{nres} } (\mathbf{r}_A) &=& \frac{ \mu_0 }{ \pi \hbar } \int_0^\infty \frac{ \xi^3d\xi }{ \xi^2 + \tilde{\omega}_{nk}^2 } \nonumber\\
&& \times \mathrm{Im} \left[ \mathbf{d}_{nk} \cdot \mathbb{G}^{(1)}(\mathbf{r}_A,\mathbf{r}_A; \mathrm{i}\xi ) \cdot \mathbf{d}_{kn} \right] \nonumber\\
&& - \frac{ \mu_0 }{ \pi \hbar } \int_0^\infty \frac{ \xi^2 \tilde{\omega}_{nk} d\xi }{ \xi^2 + \tilde{\omega}_{nk}^2 } \nonumber\\
&& \times \mathrm{Re} \left[ \mathbf{d}_{nk} \cdot \mathbb{G}^{(1)}(\mathbf{r}_A,\mathbf{r}_A; \mathrm{i}\xi ) \cdot \mathbf{d}_{kn} \right] , \; \label{Nres Freq shift r}
\end{eqnarray}
with a Green's function $\mathbb{G}^{(1)}$ with imaginary frequency $\omega\rightarrow \mathrm{i}\xi$. 

The evaluation of the poles gives the resonant contribution associated with the real-photon emission and a real-frequency expression $\tilde{\omega}_{nk}$ \cite{Fuchs-Crosse-Buhmann}
\begin{equation} \label{Res Freq shift r}
\delta\omega_{nk}^{ \mathrm{res} } (\mathbf{r}_A) =  - \frac{ \mu_0 }{ \hbar }\tilde{\omega}_{nk}^2 \mathrm{Re} \left[ \mathbf{d}_{nk} \cdot \mathbb{G}^{(1)}(\mathbf{r}_A,\mathbf{r}_A; \tilde{\omega}_{nk} ) \cdot \mathbf{d}_{kn} \right] \;.
\end{equation}

From Eq.~(\ref{Nres Freq shift r}) 
we observe that the matrix-vector product of the Green's tensor and the dipole moments is real for a reciprocal medium and therefore only the second contribution remains in this case \cite{Fuchs-Crosse-Buhmann}. Additionally, in Eq.~(\ref{Freq shift r v}) 
we appreciate another requirement for a non-zero contribution, which is that the Green's tensor of the corresponding material must have a curl different from zero. Remarkably, this last condition is satisfied for certain reciprocal media opening the gate to probe quantum friction in chiral materials for example. The case of chiral media will be analyzed in detail in the following section. 

\section{Applications} \label{APPS}

\begin{figure}[tbp]
\centering 
\includegraphics[width=8CM]{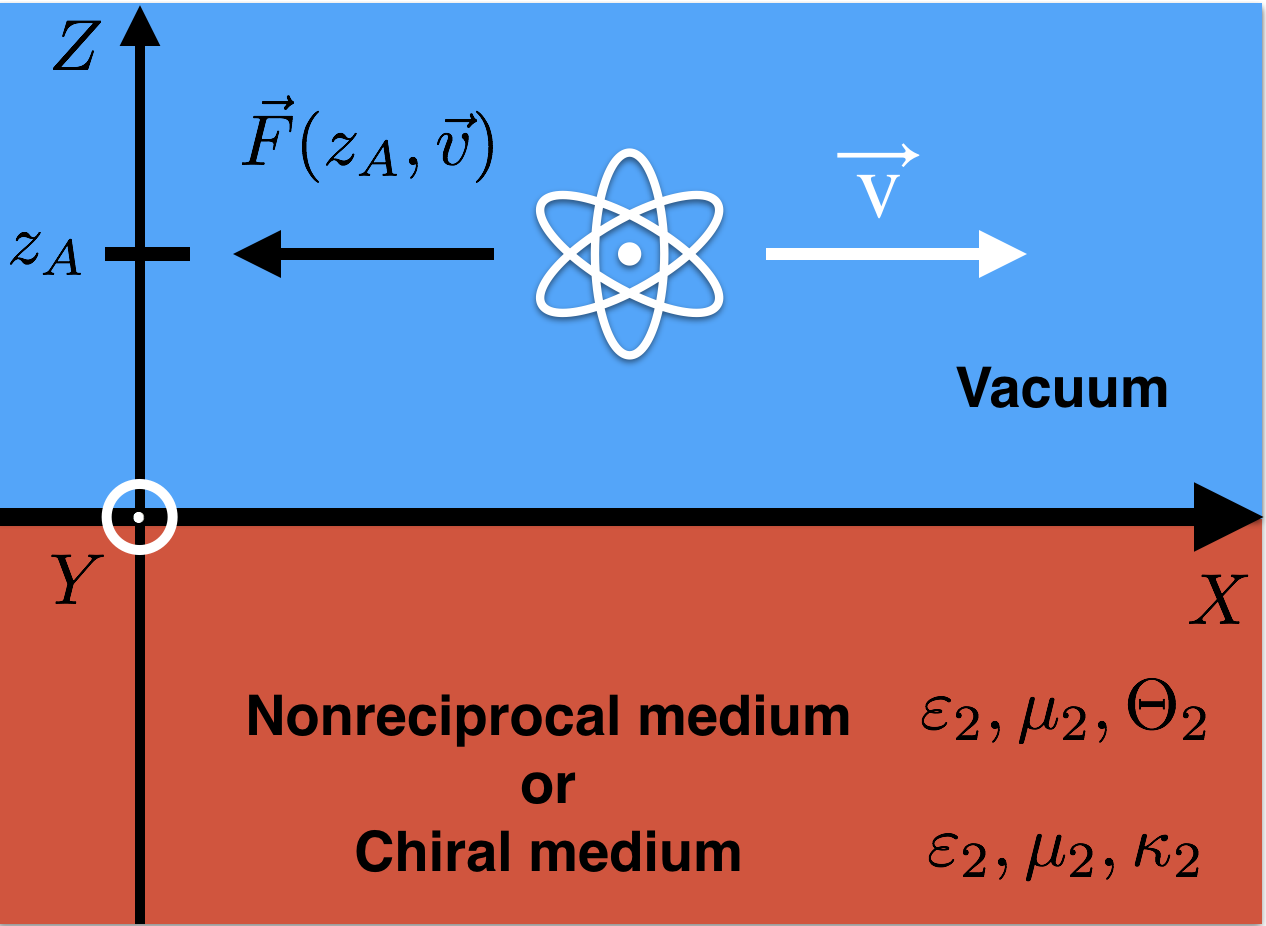} 
\caption{ Quantum friction experienced by an atom moving with constant velocity parallel to a planar interface made of vacuum and a nonreciprocal medium or of vacuum and a chiral medium. The experienced force is $\mathbf{F}(z_A, \mathbf{v})$, but is not treated in this work. Instead we study the frequency shift and decay rate induced in the atom.}
\label{SETUP}
\end{figure}

Once we have derived expressions for 
the frequency shift Eqs.~(\ref{Freq shift r v}), (\ref{Nres Freq shift r}) and (\ref{Res Freq shift r}), we proceed to apply our findings for three ideal media: a perfectly conducting mirror, a perfectly reflecting nonreciprocal mirror, and a perfectly reflecting chiral mirror. Afterward, as more realistic examples, we illustrate our results for a strong three-dimensional TI and an isotropic chiral medium. In this work, we consider the configuration depicted in Fig.~\ref{SETUP}, where we consider a bi-layer configuration with the vacuum as upper layer and the nonreciprocal medium or the chiral medium as the lower layer. For this selection of five media, we restrict ourselves to real and constant permittivity, permeability and cross-susceptibility unless the contrary is specified. 

The scattering part of the Green's tensor $\mathbb{G}^{(1)}$ of a single planar surface has the form \cite{Crosse-Fuchs-Buhmann}
\begin{eqnarray}
\mathbb{G}^{(1)}(\mathbf{r},\mathbf{r}'; \omega) &=& \frac{ \mathrm{i} }{ 8\pi^2 }\int \frac{ d^2\mathbf{k}_\parallel }{ k_\perp } \sum_{ \sigma= \mathrm{s,p} } \sum_{ \sigma' = \mathrm{s,p} } r_{ \sigma,\sigma' } \nonumber\\
&& \times \mathbf{e}_{\sigma +}\otimes\mathbf{e}_{\sigma' -} \mathrm{e}^{ \mathrm{i} \mathbf{k}_\parallel \cdot (\mathbf{r}-\mathbf{r}') } \mathrm{e}^{ \mathrm{i} k_\perp (z+z') } \;, \quad \label{GF dyadics}
\end{eqnarray}
where $\mathbf{e}_{\sigma +}$ and $\mathbf{e}_{\sigma' -}$ are the polarization unit vectors for $\sigma$- and $\sigma'$-polarized waves in the upper layer of Fig.~\ref{SETUP}. The indices $\mathrm{s}$ and $\mathrm{p}$ refer to perpendicular or parallel polarization, respectively. Here $\mathbf{k}_\parallel$ represents the parallel component of the wave vector, $k_\perp$ stands for its perpendicular component, and $z$ denotes the vertical distance to the interface. 

As noted in Section \ref{INTERNAL}, Curie’s principle \cite{Curie} dictates that to probe an atom’s sensitivity to the broken time-reversal symmetry of a nonreciprocal or the parity symmetry breaking of a chiral medium, its dipole moments must share the same symmetry violation. To be as general as possible, we choose the following dipole moments
\begin{equation}\label{Dipole}
\mathbf{d}_{nk}=\left( d_{nk}^{\,x},d_{nk}^{\,y},d_{nk}^{\,z} \right)^\top \;,\; \mathbf{d}_{kn} = \mathbf{d}_{nk}^* \;. 
\end{equation}
where $d_{nk}^{\,x},d_{nk}^{\,y},d_{nk}^{\,z}\in\mathbb{C}$ implying that such dipoles are not invariant if the direction of time is reversed $t\rightarrow- t$ as long as two of their components are not null. Actually, if $d_{nk}^{\,x}=d_{nk}/\sqrt{2}$, $d_{nk}^{\,y}=\mathrm{i}d_{nk}/\sqrt{2}$ and $d_{nk}^{\,z}=0$ give circularly polarized dipoles which were employed in the article \cite{Fuchs-Crosse-Buhmann} and their fields propagate along the $z$ direction. Additionally, due to our interest in quantum friction closer to the interface between the materials, we will take the velocity of the atom with $v_z=\mathbf{v}\cdot\mathbf{e}_z=0$, meaning that all the velocity-dependent quantities will depend only on $\mathbf{v}_\parallel=v_x\mathbf{e}_x  + v_y\mathbf{e}_y$. Also we have to consider another constriction for the dipole moments and the velocity $\mathbf{v}_\parallel\cdot\mathbf{d}_{nk}=0$, which is imposed by Maxwell equations.


\subsection{Perfectly conducting mirror}\label{Conducting mirror}
The reflective coefficients for a perfectly conducting mirror are $r_{ \mathrm{p,p} }=1$, $r_{ \mathrm{s,s} }=-1$ and $r_{ \mathrm{p,s} } = r_{ \mathrm{s,p} }=0$. This set of coefficients is obtained from the reflective coefficients for a general material in the limit $\varepsilon\rightarrow\infty$, which is explained in detail in Ref.~\cite{Fuchs-Crosse-Buhmann}.  For this mirror the Green's tensor (\ref{GF dyadics}) is only diagonal with
\begin{eqnarray}
\mathbb{G}^{(1)}_{xx} (\mathbf{r},\mathbf{r}; \omega) = \left( - \frac{ 1 }{ 8\pi z } - \frac{ \mathrm{i}\,c }{ 16\pi \omega z^2 } + \frac{ c^2 }{ 32\pi \omega^2 z^3 } \right) \mathrm{e}^{ \frac{ 2\mathrm{i} \omega z }{ c } } , \nonumber\\
\end{eqnarray}
\begin{eqnarray}
\mathbb{G}^{(1)}_{yy} (\mathbf{r},\mathbf{r}; \omega) &=& \mathbb{G}^{(1)}_{xx} (\mathbf{r},\mathbf{r}; \omega) \;, \\
\mathbb{G}^{(1)}_{zz} (\mathbf{r},\mathbf{r}; \omega) &=& \left( -\frac{ \mathrm{i}\,c }{ 8\pi\omega z^2 } + \frac{ c^2 }{ 16\pi\omega^2 z^3 } \right) \mathrm{e}^{ \frac{ 2\mathrm{i} \omega z }{ c } } \;.
\end{eqnarray}

The contributions to the frequency shift (\ref{whole shift}) for the dipole (\ref{Dipole}) in the retarded limit ($\tilde{\omega}_{nk}z_A/c\gg1$) read as follows
\begin{eqnarray}
\delta\omega^{ \mathrm{res, ret} }_{nk} (\mathbf{r}_A) &=& \frac{ \mu_0 \tilde{\omega}_{nk}^2 }{ 8 \pi \hbar } \left[ \frac{ |\mathbf{d}_{nk}^{\parallel}|^2  }{ z_A } \cos\left( \frac{ 2\tilde{\omega}_{nk}z_A }{ c } \right) \right. \nonumber\\
&& \left. - \frac{ c |d_{nk}^{\,z}|^2 }{ \tilde{\omega}_{nk} z_A^2 } \sin\left( \frac{ 2\tilde{\omega}_{nk}z_A }{ c } \right) \right]\;, \label{Shift r res ret pcm} \\
\delta\omega^{ \mathrm{nres, ret} }_{nk} (\mathbf{r}_A)
&=&  \frac{ c |\mathbf{d}_{nk}|^2 }{ 32 \pi^2 \hbar \varepsilon_0 \tilde{\omega}_{nk} z_A^4 }  \;,  \label{Shift r nres ret pcm} \\
\delta\omega^{ \mathrm{ ret} }_{nk} (\mathbf{r}_A, \mathbf{v}) &=& \frac{ \mu_0 \tilde{\omega}_{nk}^2 S_{nk}(\mathbf{v}_\parallel)}{ 4 \pi \hbar cz_A } \sin\left( \frac{ 2\tilde{\omega}_{nk}z_A }{ c } \right) \nonumber\\
&=& 0\;, \label{Shift r v ret pcm}
\end{eqnarray}
where $\mathbf{d}_{nk}^{\parallel}=d_{nk}^{\,x}\mathbf{e}_x  +  d_{nk}^{\,y}\mathbf{e}_y $ and $z_A$ is the distance of the atom to the interface. For the nonretarded limit ($\tilde{\omega}_{nk}z_A/c\ll1$)
, we find
\begin{eqnarray}
\delta\omega^{ \mathrm{res, nret} }_{nk} (\mathbf{r}_A) &=& - \frac{ 1 }{ 16 \pi \hbar \varepsilon_0 z_A^3 } \left[ \frac{1}{2}|\mathbf{d}_{nk}^{\parallel}|^2 + |d_{nk}^{\,z}|^2 \right]\;,\quad\quad \label{Shift r res nret pcm} \\
\delta\omega^{ \mathrm{nres, nret} }_{nk} (\mathbf{r}_A) &=& \frac{ 1 }{ 32 \pi^2 \hbar \varepsilon_0 z_A^3 } \left[ \frac{1}{2}|\mathbf{d}_{nk}^{\parallel}|^2 + |d_{nk}^{\,z}|^2 \right] \;, \label{Shift r nres nret pcm} \\
\delta\omega^{ \mathrm{ nret} }_{nk} (\mathbf{r}_A, \mathbf{v}) &=&  \frac{ \mu_0 \tilde{\omega}_{nk} S_{nk}(\mathbf{v}_\parallel) }{ 8 \pi \hbar   z_A^2 } = 0  \;. \label{Shift r v nret pcm} 
\end{eqnarray}

We observe that the relevant vectorial structure of the frequency shifts (\ref{Shift r res ret pcm}), (\ref{Shift r nres ret pcm}), 
(\ref{Shift r res nret pcm}) and (\ref{Shift r nres nret pcm}) 
depends only on the moduli of the dipole moments components meaning that they will contribute to the whole quantities given in Eqs.~(\ref{whole rate}) and (\ref{whole shift}) no matter whether the dipole moments are real or complex. This reflects that the perfectly conducting mirror and its corresponding dipole moments do not exhibit parity breaking or time-reversal symmetry breaking, i.e., they are $P$--even and $T$--even. In Table \ref{Tab-Symmetries} we summarize the behavior that a perfectly conducting mirror and its dipole moments $\mathbf{d}_{nk}$ must have under $P$ and $T$ transformations. Meanwhile, the shift (\ref{Shift r v ret pcm}) and (\ref{Shift r v nret pcm}) 
are governed by 
\begin{equation}\label{OLS}
S_{nk}(\mathbf{v}_\parallel) = \mathrm{Im}\left[ \left( \mathbf{d}_{nk}^* \cdot \mathbf{e}_z \right) \left( \mathbf{d}_{nk} \cdot \mathbf{v}_\parallel \right) \right] =0 \;,
\end{equation}
which has the same units as the optical rotatory strength found in chiral molecules \cite{Butcher-Buhmann-Scheel, Molecular QED} but is a vanishing quantity because Maxwell equations require that $\mathbf{v}_\parallel\cdot\mathbf{d}_{nk}=0$ so the $P$--even and $T$--even behavior of this material is respected. Lastly, we remark that due to our approximation at first order in $v/c$, the sign of the velocity could lead to an increase or decrease in the velocity-dependent frequency shifts.

\begin{table*}[ht]
\centering
\begin{tabular}{||c c c c c c c ||} 
\hline
Medium & $P$ & $T$ & Atom & $P$ & $T$ & $\delta\omega_{nk}$ \\
\hline
Perfectly conducting mirror & + & + & $|\mathbf{d}_{nk}|^2$ & + & + & + \\
Nonreciprocal mirror & $-$ & $-$ & $\mathcal{B}_z,T_{nk}(\mathbf{v}_\parallel)$ & $-$ & $-$ & + \\
3D topological insulator & $-$ & $-$ & $\mathcal{B}_z,T_{nk}(\mathbf{v}_\parallel)$ & $-$ & $-$ & + \\
Chiral mirror & $-$ & + & $R_{nk}(\mathbf{v}_\parallel)$ & $-$ & + & + \\
Isotropic chiral medium & $-$ & + & $R_{nk}(\mathbf{v}_\parallel)$ & $-$ & + & + \\
\hline
\end{tabular}
\caption{ Behavior under parity $P$ and time-reversal $T$ of the five media considered in this work and their atomic electric dipole $\mathbf{d}_{nk}$ and its corresponding medium. The second and third columns display the 
behavior under $P$ and $T$ of the five  materials. Each medium is coupled to an atomic response whose behavior under $P$ and $T$ is shown in the fourth and fifth columns. The last column shows that the coupling of each material with its atomic response gives a $P$-- and $T$-- even frequency shift. $\mathcal{B}_z$ is given by Eq.~(\ref{d x d* ez}), $T_{nk}(\mathbf{v}_\parallel)$ is defined in Eq.~(\ref{nonreciprocal ORS}) and  $R_{nk}(\mathbf{v}_\parallel)$ in Eq.~(\ref{ORS}).}
\label{Tab-Symmetries}
\end{table*}


\subsection{Perfectly reflecting nonreciprocal mirror}\label{Nonreciprocal mirror}

This mirror is characterized by vanishing polarization preserving reflection coefficients for incoming perpendicular or parallel polarization and outgoing perpendicular or parallel polarization $r_{ \mathrm{s,s} } = r_{ \mathrm{p,p} }=0$ and therefore the mixing terms $r_{ \mathrm{p,s} }$ and $r_{ \mathrm{s,p} }$ can be chosen to be either $\pm 1$ \cite{Fuchs-Crosse-Buhmann,CPT-article}. We will restrict ourselves to the case $r_{ \mathrm{p,s} } = r_{ \mathrm{s,p} }=-1$. Under these conditions the Green's tensor (\ref{GF dyadics}) results antisymmetric $\mathbb{G}^{(1) \top }(\mathbf{r},\mathbf{r}'; \omega)=-\mathbb{G}^{(1)}(\mathbf{r}',\mathbf{r}; \omega)$ whose only nonzero components are \cite{Fuchs-Crosse-Buhmann}
\begin{eqnarray}
\mathbb{G}^{(1)}_{xy} (\mathbf{r},\mathbf{r}; \omega) &=&  \left( - \frac{ 1 }{ 8\pi z } - \frac{ \mathrm{i}\,c }{ 16\pi \omega z^2 } \right) \mathrm{e}^{ \frac{ 2\mathrm{i} \omega z }{ c } } \;, \\
\mathbb{G}^{(1)}_{yx} (\mathbf{r},\mathbf{r}; \omega) &=& -\mathbb{G}^{(1)}_{xy} (\mathbf{r},\mathbf{r}; \omega) \;.
\end{eqnarray}

The contributions to the frequency shift (\ref{whole shift}) for the dipole (\ref{Dipole}) in the retarded limit ($\tilde{\omega}_{nk}z_A/c\gg1$) read as follows
\begin{eqnarray}
\delta\omega^{ \mathrm{res, ret} }_{nk} (\mathbf{r}_A) &=& -\frac{ \mu_0 \tilde{\omega}_{nk}^2 \mathcal{B}_z }{ 4 \pi \hbar z_A } 
%
\sin\left( \frac{ 2\tilde{\omega}_{nk}z_A }{ c } \right) \;, \label{Shift r res ret prnm} \\
\delta\omega^{ \mathrm{nres, ret} }_{nk} (\mathbf{r}_A) &=& - \frac{ 3 c^2 \mathcal{B}_z }{ 32 \pi^2 \hbar \varepsilon_0 \tilde{\omega}_{nk}^2 z_A^5 } 
\;, \label{Shift r nres ret prnm} \\
\delta\omega^{ \mathrm{ ret} }_{nk} (\mathbf{r}_A, \mathbf{v}) &=& \frac{ \mu_0 \tilde{\omega}_{nk}^2 }{ 4 \pi \hbar cz_A } T_{nk}(\mathbf{v}_\parallel) 
%
\cos\left( \frac{ 2\tilde{\omega}_{nk}z_A }{ c } \right),\;\; \label{Shift r v ret prnm}  
\end{eqnarray}
where the dominant contribution to the frequency shift (\ref{Shift r v ret prnm}) comes from the electric interaction $\mathbf{d}_{mn} \cdot \mathbf{ \hat{E} }(\mathbf{r}_A)$. For the nonretarded limit ($\tilde{\omega}_{nk}z_A/c\ll1$), we find
\begin{eqnarray}
\delta\omega^{ \mathrm{res, nret} }_{nk} (\mathbf{r}_A) &=& -\frac{ \mu_0 c\, \tilde{\omega}_{nk} }{ 8 \pi \hbar z_A^2 } \mathcal{B}_z 
\;, \label{Shift r res nret prnm} \\ 
\delta\omega^{ \mathrm{nres, nret} }_{nk} (\mathbf{r}_A) &=& - \frac{ \mathcal{B}_z }{ 16 \pi^2 \hbar \varepsilon_0 z_A^3 }  \;, \label{Shift r nres nret prnm} \label{eqn:frequshiftbarenonres} \\
\delta\omega^{ \mathrm{ nret} }_{nk} (\mathbf{r}_A, \mathbf{v}) &=& -\frac{ T_{nk}(\mathbf{v}_\parallel) }{ 16 \pi \hbar \varepsilon_0 c z_A^3 } 
%
\;. \label{Shift r v nret prnm}
\end{eqnarray}

For this mirror, we identify that the relevant vectorial structure of Eqs.~(\ref{Shift r res ret prnm}), (\ref{Shift r nres ret prnm}), 
(\ref{Shift r res nret prnm}) and (\ref{Shift r nres nret prnm}), 
depends on 
\begin{equation}\label{d x d* ez}
\mathcal{B}_z=\mathrm{Im}(d_{nk}^{\,x}d_{nk}^{\,y\,*}) =  \frac{1}{ 2 \mathrm{i} } \left( \mathbf{d}_{nk} \times \mathbf{d}_{nk}^* \right) \cdot \mathbf{e}_z \;,
\end{equation}
which is $T$-odd and  shows that $\mathbf{d}_{nk}$, $\mathbf{d}_{nk}^*$ and $\mathbf{e}_z$ must not be coplanar. In this way, the time-reversal symmetry breaking for such position-dependent quantities becomes explicitly manifest in comparison with the previous study \cite{Fuchs-Crosse-Buhmann} because only circular or elliptical polarizations parallel to the $xy$ plane will give a non-zero contribution. In other words, linear polarizations are ruled out because they give $\mathbf{d}_{nk} \times \mathbf{d}_{nk}^*=\mathbf{0}$ as a consequence of their invariance under time-reversal symmetry and colinearity. Important to highlight is the cross product of Eq.~(\ref{d x d* ez}), which in the framework of chiral media can be understood as a geometric magnetic field $\mathcal{B}$ associated with photoionization of chiral molecules \cite{Ayuso-Ordonez-Smirnova,Ordonez-Smirnova} and whose $z$ component is the only relevant here. This offers another first connection between the physics of chirality and nonreciprocity. \\

On the other hand, Eqs.~(\ref{Shift r v ret prnm}) and (\ref{Shift r v nret prnm}) 
share the same vectorial structure for the atomic response
\begin{equation}
T_{nk}(\mathbf{v}_\parallel) = \mathrm{Re}\left\{ \left( \mathbf{d}_{nk}^* \cdot \mathbf{e}_z \right) \left[ \left( \mathbf{v}_\parallel \times \mathbf{d}_{nk} \right) \cdot \mathbf{e}_z \right] \right\} \;,
\end{equation}
which is $T$-odd and does not vanish if $\mathbf{v}_\parallel$, $\mathbf{d}_{nk}$ and $\mathbf{e}_z$ are not coplanar and if $d_{nk}^{\,z}\neq 0$. These conditions imply the necessity of a superposition of linear polarizations to obtain a non-zero contribution 
for the velocity-dependent quantities opposed to what the position-dependent ones required via $\mathcal{B}_z$ (\ref{d x d* ez}). Besides $d_{nk}^{\,z}\neq 0$ this vectorial structure needs either $d_{nk}^{\,x}\neq 0$ or $d_{nk}^{\,y}\neq 0$, which results in the surprising possibility to turn off only the position-dependent of Eqs.~(\ref{Shift r res ret prnm}), (\ref{Shift r nres ret prnm}), (\ref{Shift r res nret prnm}) and (\ref{Shift r nres nret prnm}). Namely, if we choose the polarized dipole in the $xz$ plane $\mathbf{d}_{nk}=d_{nk}/\sqrt{2}\left(0,1,1\right)^\top$, for which $d_{nk}^{\,x\,*}=0$, then Eq.~(\ref{d x d* ez}) is zero but $\mathrm{Re}\left\{ \left( \mathbf{d}_{nk}^* \cdot \mathbf{e}_z \right) \left[ \left( \mathbf{v}_\parallel \times \mathbf{d}_{nk} \right) \cdot \mathbf{e}_z \right] \right\}\neq 0$ leading to non-zero velocity-dependent quantities. 

Given the significance of this vectorial structure, as a next step, we examine its origin. In this way, we recall the interaction Hamiltonian and regard only the R\"ontgen interaction, which can be rewritten as shown:
\begin{equation}\label{Roentgen}
\hat{ H }_{ \text{R\"ontgen} }= \mathbf{v} \cdot \mathbf{ \hat{d} } \times \mathbf{ \hat{B} }(\mathbf{r}_A) = - \mathbf{ \hat{m} } \cdot \mathbf{ \hat{B} }(\mathbf{r}_A) \;,
\end{equation}
where
\begin{equation} \label{eff magnetic moment}
\mathbf{ \hat{m} } = - \mathbf{v} \times \mathbf{ \hat{d} } \;.
\end{equation}
This means that an effective magnetic moment is induced. In this way, the vectorial structure for the velocity-dependent quantities takes the following form 
\begin{eqnarray}
\mathrm{Re}\left\{ \left( \mathbf{d}_{nk}^* \cdot \mathbf{e}_z \right) \left[ \left( \mathbf{v}_\parallel \times \mathbf{d}_{nk} \right) \cdot \mathbf{e}_z \right] \right\} &=& -\mathrm{Re}\left( d_{nk}^{\,z\,*} m_{nk}^{\,z} \right) \;, \nonumber\\
&=& T_{nk}(\mathbf{v}_\parallel) \; , \label{nonreciprocal ORS} 
\end{eqnarray}
which resembles the optical rotatory strength $R_{nk}=-\mathrm{Im}\left( d_{nk}^{\,z\,*}\cdot m_{nk}\right)$ of chiral molecules \cite{Butcher-Buhmann-Scheel, Molecular QED} but this one needs the real part implying the necessity of real dipoles. The form of $T_{nk}(\mathbf{v}_\parallel)$ is in agreement with the Casimir--Polder potential perspective studied on Ref.~\cite{CPT-article}, where for nonreciprocal media it is required that  $\mathbf{d}_{nk}\otimes\mathbf{m}_{kn}$ must be real as part of the crossed polarisability. In order to distinguish between the nonreciprocal and the ordinary rotatory strangths, we will name $T_{nk}(\mathbf{v}_\parallel)$ as nonreciprocal optical rotatory strength. This constitutes the second connection found in this work between the physics of chirality and nonreciprocity. In Table \ref{Tab-Symmetries} we summarize the behavior that a perfectly reflecting nonreciprocal mirror and its dipole moments $\mathbf{d}_{nk}$ must have under $P$ and $T$ transformations. %

Our two final remarks for this section are as follows. First, without velocity the nonreciprocal optical rotatory strength $T_{nk}$ is zero, thus there is no magnetic moment so the whole velocity-dependent effect is zero. And second, due to our approximation at first order in $v/c$, the sign of the velocity could lead to an increase or decrease in the velocity-dependent frequency shifts 
as happened to the perfectly conducting mirror discussed in the previous subsection.


\subsection{3D topological insulator}\label{3D TI}

For these materials, the electric and magnetic field are connected with the displacement field $\hat{\mathbf{D}}\left(\mathbf{r};\omega\right)$ and the magnetic field $\hat{\mathbf{H}}\left(\mathbf{r};\omega\right)$ via
\begin{eqnarray}
\hat{\mathbf{D}}\left(\mathbf{r};\omega\right) &=& \varepsilon_0\varepsilon\left(\mathbf{r};\omega\right) \hat{\mathbf{E}}\left(\mathbf{r};\omega\right) + \frac{ \alpha \Theta\left(\mathbf{r};\omega\right) }{ \pi\mu_0 c }  \hat{\mathbf{B}}\left(\mathbf{r};\omega\right) \nonumber\\
&& + \hat{\mathbf{P}}_N\left(\mathbf{r};\omega\right) \;, \label{Constitutive 1}
\end{eqnarray}
\begin{eqnarray}
\hat{\mathbf{H}}\left(\mathbf{r};\omega\right) &=& \frac{ \hat{\mathbf{B}}\left(\mathbf{r};\omega\right) }{ \mu_0\mu\left(\mathbf{r};\omega\right) } -  \frac{ \alpha \Theta\left(\mathbf{r};\omega\right) }{ \pi\mu_0 c }  \hat{\mathbf{E}}\left(\mathbf{r};\omega\right) \nonumber\\
&& - \hat{\mathbf{M}}_N\left(\mathbf{r};\omega\right) \;, \label{Constitutive 2}
\end{eqnarray}
where $\alpha$ is the fine structure constant and $\varepsilon\left(\mathbf{r};\omega\right)$, $\mu\left(\mathbf{r};\omega\right)$, and $\Theta\left(\mathbf{r};\omega\right)$ are the dielectric permittivity, magnetic permeability, and axion coupling, respectively. Here the $\hat{\mathbf{P}}_N\left(\mathbf{r};\omega\right)$ and $\hat{\mathbf{M}}_N\left(\mathbf{r};\omega\right)$ terms are the noise polarization and magnetization, respectively. These are Langevin noise terms that model absorption within the material \cite{Acta}. The topological nature of these materials provides a distinction between strong TIs (odd number of surface states per surface) from weak ones (even number on all but one surface) \cite{TIs Prediction 1, Morgenstern review}.  The axion coupling $\Theta\left(\mathbf{r};\omega\right)$ takes even multiples of $\pi$ in weak three-dimensional TIs and odd multiples of $\pi$ in strong three-dimensional TIs, with the magnitude and sign of the multiple given by the strength and direction of the time-symmetry-breaking perturbation. Although, the weak class of TIs is under active research it provides more complications because lattice translation symmetry together with time-reversal symmetry protect the conducting states at the boundaries of a weak 3D TI, which requires to take into account also disorder and anisotropies Ref.~\cite{TIs Prediction 1,WTIs-Nature,WTIs}. For these reasons, in this work we will only consider the strong class of them. When $\Theta$ is restricted to a nondynamical field, the theory is usually referred as $\Theta$-Electrodynamics \cite{OJF-LFU-ORT, AMR-LFU-MC 1, AMR-LFU-MC 2}, which has the necessary topological features to model the electromagnetic behavior of strong 3D TIs \cite{Qi PRB} and also naturally existing magnetoelectric media \cite{Hehl,Landau-Lifshitz} such as Cr$_2$O$_3$ with $\Theta\simeq\pi/36$ \cite{Wu,Hehl-Obukov-Rivera}. 

The modified Fresnel reflective coefficients for nonreciprocal media $r_{ \sigma,\sigma' }$ involved in Eq.~(\ref{GF dyadics}) can be derived from the application of the boundary conditions to above constitutive relations. These  reflective coefficients can be found in Ref.~\cite{Crosse-Fuchs-Buhmann} in its general form. However, we will particularize them for the case of an interface constituted by vacuum in the upper layer of Fig.~\ref{SETUP}  and a strong three-dimensional TI with real permittivity in the lower layer as shown
\begin{eqnarray} 
r_{ \mathrm{s,s} } &=& \frac{(\mu_2 k_1^\perp - k_2^\perp) \mu_2 (  \varepsilon_2 k_1^\perp + k_2^\perp) - \Delta^{2}k_1^\perp k_2^\perp }{(\mu_2 k_1^\perp + k_2^\perp) \mu_2 (  \varepsilon_2 k_1^\perp + k_2^\perp) + \Delta^{2}k_1^\perp k_2^\perp } , \label{Fresnel 1} \\
r_{ \mathrm{s,p} } &=& \frac{ -2\mu_2 k_1^\perp k_2^\perp \Delta}{(\mu_2 k_1^\perp + k_2^\perp) \mu_2 (  \varepsilon_2 k_1^\perp + k_2^\perp) + \Delta^{2}k_1^\perp k_2^\perp } , \label{Fresnel 2} \\
r_{ \mathrm{p,p} } &=& \frac{(  \varepsilon_2 k_1^\perp - k_2^\perp)\mu_2(\mu_2 k_1^\perp + k_2^\perp) + \Delta^{2}k_1^\perp k_2^\perp }{(\mu_2 k_1^\perp + k_2^\perp) \mu_2 (  \varepsilon_2 k_1^\perp + k_2^\perp) + \Delta^{2}k_1^\perp k_2^\perp } , \label{Fresnel 3} \\
r_{ \mathrm{p,s} } &=& r_{ \mathrm{s,p} } \;, \label{Fresnel 4} 
\end{eqnarray}
where the subscript 1 denotes the corresponding quantities in the vacuum and conversely the subscript 2 for the strong 3D TI. The perpendicular part of the wave vector is denoted by $k_1^\perp$ and $k_2^\perp$ for the corresponding medium. For the general case when two semi-infinite magnetoelectric media with different axion couplings constitute the planar interface, the topological parameter $\Delta$ takes the following form \cite{Crosse-Fuchs-Buhmann} 
\begin{equation}
\Delta=\alpha\mu_1\mu_2(\Theta_2-\Theta_1)/\pi \;.
\end{equation}
Nevertheless, for the selected interface the strong 3D TI with $\Theta_1=\pi$ and vacuum with $\Theta_2=0$, the topological parameter takes the next form 
\begin{equation}\label{DELTA} 
\Delta=\alpha(2m+1), 
\end{equation}
where $m$ is an integer depending on the details of the time-reversal-symmetry breaking at the interface.


The contributions to the frequency shift (\ref{whole shift}) for the dipole (\ref{Dipole}) in the retarded limit ($\tilde{\omega}_{nk}z_A/c\gg1$) read as follows
\begin{eqnarray}
\delta\omega^{ \mathrm{res, ret} }_{nk} (\mathbf{r}_A) &=& \frac{ \mu_0 \tilde{\omega}_{nk}^2 }{ 4 \pi \hbar } \left[ \mathcal{B}_z \frac{ r^{ \mathrm{ret} }_{ \mathrm{s,p} } }{ z_A } \sin\left( \frac{ 2\tilde{\omega}_{nk}z_A }{ c } \right) \right. \nonumber \\
&& + |\mathbf{d}_{nk}^{\parallel}|^2 \frac{ r^{ \mathrm{ret} }_{ \mathrm{p,p} } }{ 2z_A } \cos\left( \frac{ 2\tilde{\omega}_{nk}z_A }{ c } \right)  \nonumber\\
&&  \left. - \frac{ c\, r^{ \mathrm{ret} }_{ \mathrm{ p,p} } |d_{nk}^{\,z}|^2 }{ 2 \tilde{\omega}_{nk} z_A^2 } \sin\left( \frac{ 2\tilde{\omega}_{nk}z_A }{ c } \right) \right] \;, \label{Shift r res ret TI} \\
\delta\omega^{ \mathrm{nres, ret} }_{nk} (\mathbf{r}_A) &=& \frac{ c }{ 32 \pi^2 \hbar \varepsilon_0 \tilde{\omega}_{nk} z_A^4 } \left[ \mathcal{B}_z \frac{ 3c\, r^{ \mathrm{ret} }_{ \mathrm{s,p} } }{ \tilde{\omega}_{nk}z_A } \right. \nonumber\\
&& \left. + r^{ \mathrm{ret} }_{ \mathrm{ p,p} } |\mathbf{d}_{nk}|^2 \right] \;, \label{Shift r nres ret TI} \\
\delta\omega^{ \mathrm{ ret} }_{nk} (\mathbf{r}_A, \mathbf{v}) &=& \frac{ \mu_0 \tilde{\omega}_{nk}^2 }{ 4 \pi \hbar cz_A }  r^{ \mathrm{ret} }_{ \mathrm{s,p} } T_{nk}(\mathbf{v}_\parallel) \cos\left( \frac{ 2\tilde{\omega}_{nk}z_A }{ c } \right) 
%
\;, \label{Shift r v ret TI} \nonumber\\
\end{eqnarray}
where $S_{nk}(\mathbf{v}_\parallel)=0$ from Eq.~(\ref{OLS}) was used and the nonreciprocal optical rotatory strength $T_{nk}(\mathbf{v}_\parallel)$ was defined in Eq.~(\ref{nonreciprocal ORS}), and both of them will be subsequently employed. The retarded limit of the Fresnel coefficients are given in Appendix \ref{A}. Once again, the dominant contribution to the frequency shift (\ref{Shift r v ret TI}) arises from the electric interaction $\mathbf{d}_{mn} \cdot \mathbf{ \hat{E} }(\mathbf{r}_A)$ similar to the frequency shift of the  perfectly reflecting nonreciprocal mirror (\ref{Shift r v ret prnm}). For the nonretarded limit ($\tilde{\omega}_{nk}z_A/c\ll1$), we obtain
\begin{eqnarray}
&&\delta\omega^{ \mathrm{res, nret} }_{nk} (\mathbf{r}_A) = \frac{ \mu_0 \tilde{\omega}_{nk}^2  }{ 8 \pi \hbar } \left[ \mathcal{B}_z \frac{ r^{ \mathrm{nret} }_{ \mathrm{s,p} } c }{ \tilde{\omega}_{nk} z_A^2 } \right. \nonumber\\
&& \left. + |\mathbf{d}_{nk}^{\parallel}|^2  \left( \frac{ r^{ \mathrm{nret} }_{ \mathrm{s,s} } }{ 2z_A } + \frac{ r^{ \mathrm{nret} }_{ \mathrm{p,p} } c^2 }{ \tilde{\omega}_{nk}^2 z_A^3 } \right)  + \frac{ c^2 r^{ \mathrm{nret} }_{ \mathrm{ p,p} }  |d_{nk}^{\,z}|^2 }{ 2 \tilde{\omega}_{nk}^2 z_A^3 } \right] , \label{Shift r res nret TI}
\end{eqnarray}
\begin{eqnarray}
\delta\omega^{ \mathrm{nres, nret} }_{nk} (\mathbf{r}_A) &=& \frac{ 1 }{ 16 \pi^2 \hbar \varepsilon_0 z_A^3 } \left[ \mathcal{B}_z r^{ \mathrm{nret} }_{ \mathrm{s,p} }  \right.  \nonumber\\
&& \left. + r^{ \mathrm{nret} }_{ \mathrm{p,p} } \pi \left( \frac{ |\mathbf{d}_{nk}^{\parallel}|^2 }{ 4 } + \frac{ |d_{nk}^{\,z}|^2 }{ 2 }\right)  \right], \label{Shift r nres nret TI} \\
\delta\omega^{ \mathrm{ nret} }_{nk} (\mathbf{r}_A, \mathbf{v}) &=& \frac{ r^{ \mathrm{nret} }_{ \mathrm{s,p} } T_{nk}(\mathbf{v}_\parallel) }{ 16 \pi \hbar \varepsilon_0 c z_A^3 }  \;,
%
%
%
\end{eqnarray}
where the nonretarded limit of the Fresnel coefficients are given in Appendix \ref{A}. 

For the strong 3D TI, we observe that the resulting frequency shifts 
are a superposition of the corresponding quantities of the perfectly conducting mirror of Sec. \ref{Conducting mirror} and the perfectly reflecting nonreciprocal mirror of Sec. \ref{Nonreciprocal mirror} determined by the Fresnel coefficients $r_{ \sigma,\sigma' }$. As a result, all the previous discussions and mainly the conditions over $\mathbf{v}_\parallel$, $\mathbf{d}_{nk}$ and $\mathbf{e}_z$ discussed previously apply. Moreover, the reflective coefficients are functions of the permittivity $\varepsilon$ and the topological parameter $\Delta$ which could help to modulate the behavior of the spectroscopical quantities. For instance, we could increase $\Delta$ by substituting the strong 3D TI by a magnetoelectric like TbPO$_4$ with $\varepsilon_2=3.4969$, $\mu_2=1$, and $\Delta=0.22$ \cite{TbPO4,Rivera}. Similarly to the sign of the velocity discussed for the perfectly conducting and the nonreciprocal mirror, the topological parameter $\Delta$ can acquire a different sign for certain strong 3D TIs. This can be easily seen from the crossed reflective coefficients in Appendix \ref{A} which are proportional to $\Delta$ in the retarded and the nonretarded limits. This would lead to an increase or decrease in the corresponding contribution. However, the whole frequency shift is $T$--even once $T$ is applied to $T_{nk}(\mathbf{v}_\parallel)$. In Table \ref{Tab-Symmetries} we resume the behavior that a strong 3D TI and its dipole moments $\mathbf{d}_{nk}$ must have under $P$ and $T$ transformations, where both atom and medium are falsely chiral. Furthermore, due to the internal connection between the frequency shift and the Casimir--Polder force, one can also be able to switch from an attractive to a repulsive force between atom and medium. Notwithstanding this possibility, an extremely large axion coupling will approximate these results to those of the perfectly conducting mirror because in that limit $r_{ \mathrm{p,s} } = r_{ \mathrm{s,p} }\rightarrow 0$, $r_{ \mathrm{s,s} }\rightarrow -1$ and $r_{ \mathrm{p,p} }\rightarrow 1$ as discussed in Ref.~\cite{Fuchs-Crosse-Buhmann}.


\subsection{Perfectly reflecting chiral mirror}\label{Chiral mirror}

Now we proceed towards chiral media. Although this kind of media are reciprocal meaning that they verify the Lorentz reciprocity principle for the Green's tensor given by Eq.~(\ref{Lorentz reciprocity}) \cite{Ali,Butcher-Buhmann-Scheel,Caloz-Sihvola}, our result for the velocity-dependent frequency shift (\ref{Freq shift r v}) 
shows that the curl of the Green's tensor must not be zero. This feature is indeed satisfied by chiral media \cite{Butcher-Buhmann-Scheel} justifying such choice. At this point, we note that the application of our results for chiral media will only require the use of Eq.~(\ref{Lorentz reciprocity})  in all the required formulae, which means that chiral media are a particular case of nonreciprocal media. The resulting expressions are
\begin{equation}
\begin{aligned}
&\delta\omega_{nk} (\mathbf{r}_A, \mathbf{v}) = \\
& \frac{ \mu_0 }{ \hbar } \left\{ \omega^2 \mathrm{Im}\left[ \mathbf{d}_{kn}\cdot ( \mathbf{v} \cdot \nabla' ) \mathbb{G}^{(1)}(\mathbf{r}_A,\mathbf{r}_A;\omega) \cdot \mathbf{d}_{nk} \right]  \right\}'_{\omega=\tilde{\omega}_{nk}} \\
& - \frac{ \mu_0 }{ \hbar } \tilde{\omega}_{nk} \mathrm{Im} \left\{ \mathbf{v} \cdot \mathbf{d}_{kn} \times   \left[ \nabla \times \mathbb{G}^{(1)}(\mathbf{r}_A,\mathbf{r}_A;\tilde{\omega}_{nk}) \right] \cdot \mathbf{d}_{nk} \right\}   \\
& + \frac{ \mu_0 }{ \hbar } \tilde{\omega}_{nk} \mathrm{Im} \left\{ \mathbf{v} \cdot \mathbf{d}_{nk} \times   \left[ \nabla \times \mathbb{G}^{(1)}(\mathbf{r}_A,\mathbf{r}_A;\tilde{\omega}_{nk}) \right] \cdot \mathbf{d}_{kn} \right\} \;, \label{Chiral Freq shift r v} 
\end{aligned}
\end{equation}
and the quantities given by Eqs.~
(\ref{Nres Freq shift r}) and (\ref{Res Freq shift r}) remain the same.

Turning to the perfectly reflecting chiral mirror, which differs from the perfectly conducting mirror in Sec. \ref{Conducting mirror} and the perfectly reflecting nonreciprocal mirror in Sec. \ref{Nonreciprocal mirror}, it has the mixing terms $r_{ \mathrm{p,s} }=-r_{ \mathrm{s,p} }=\pm\,\mathrm{i}$ and therefore $r_{ \mathrm{s,s} } = r_{ \mathrm{p,p} }=0$ \cite{Buhmann-Singer}. Here we adopt the convention that a left-handed mirror or a mirror of positive chirality rotates the polarization of an electromagnetic wave in a counterclockwise direction upon reflection when traveling alongside the wave defining the reflection coefficients $r_{ \mathrm{p,s} }=-r_{ \mathrm{s,p} }=\mathrm{i}$, whereas a right-handed mirror of negative chirality rotates it in a clockwise direction specified by the $r_{ \mathrm{p,s} }=-r_{ \mathrm{s,p} }=-\mathrm{i}$. 
Under these conditions the Green's tensor (\ref{GF dyadics}) results 
\cite{Butcher-Buhmann-Scheel}
%
%
%
\begin{eqnarray}
\mathbb{G}^{(1)}(\mathbf{r},\mathbf{r}'; \omega) &=& \frac{ \mathrm{i} c }{ 8\pi^2 \omega }\int \frac{ d^2\mathbf{k}_\parallel }{ k_\perp } \mathrm{e}^{ \mathrm{i} \mathbf{k}_\parallel \cdot (\mathbf{r}-\mathbf{r}') } \mathrm{e}^{ \mathrm{i} k_\perp (z+z') } \nonumber\\
&& \times \left(r_{ \mathrm{s,p}}\mathbf{e}_{\mathrm{s}}\otimes\mathbf{e}_{\mathrm{p} -} +r_{ \mathrm{p,s} }\mathbf{e}_{\mathrm{p} +}\otimes\mathbf{e}_{\mathrm{s}} \right) \;. \quad 
\end{eqnarray}

The contributions to the frequency shift (\ref{whole shift}) for the dipole (\ref{Dipole}) in the retarded limit ($\tilde{\omega}_{nk}z_A/c\gg1$) read as follows
\begin{eqnarray}
&&\delta\omega^{ \mathrm{res, ret} }_{nk} (\mathbf{r}_A) = 0\;, \label{Shift r res ret prcm} \\
&&\delta\omega^{ \mathrm{nres, ret} }_{nk} (\mathbf{r}_A) = 0, \label{Shift r nres ret prcm} \\
&&\delta\omega^{ \mathrm{ ret} }_{nk} (\mathbf{r}_A, \mathbf{v}) = \pm \frac{ \tilde{\omega}_{nk}^2 R_{nk}(\mathbf{v}_\parallel) }{ 4 \pi \hbar \varepsilon_0 c^2 z_A }  \cos\left( \frac{ 2\tilde{\omega}_{nk}z_A }{ c } \right) , \label{Shift r v ret prcm}
\end{eqnarray}
%
where the dominant contribution to the frequency shift (\ref{Shift r v ret prcm}) arises from the electric interaction $\mathbf{d}_{mn} \cdot \mathbf{ \hat{E} }(\mathbf{r}_A)$, as in the  frequency shifts of the perfectly reflecting nonreciprocal mirror (\ref{Shift r v ret prnm}) and the 3D TI (\ref{Shift r v ret TI}). For the nonretarded limit ($\tilde{\omega}_{nk}z_A/c\ll1$), we find
\begin{eqnarray}
\delta\omega^{ \mathrm{res, nret} }_{nk} (\mathbf{r}_A) &=& 0\;, \label{Shift r res nret prcm} \\
\delta\omega^{ \mathrm{nres, nret} }_{nk} (\mathbf{r}_A) &=& 0\;, \label{Shift r nres nret prcm} \\
\delta\omega^{ \mathrm{ nret} }_{nk} (\mathbf{r}_A, \mathbf{v}) &=& \mp \frac{ R_{nk}(\mathbf{v}_\parallel) }{ 8 \pi \hbar \varepsilon_0 c z_A^3 }  \;. \label{Shift r v nret prcm}
\end{eqnarray}

For this mirror, we observe that all the position-dependent spectroscopical quantities are zero. This is explained by means of the Curie's principle \cite{Curie} mentioned in Sec. \ref{INTERNAL}, which applied for this situation establishes the necessity of a chiral source to probe the chirality of the mirror. This means that we need not only an electric dipole but also a magnetic dipole, for which our effective magnetic dipole of Eq.~(\ref{eff magnetic moment}) does not work because velocities are not taken into account by 
the position-dependent frequency shifts (\ref{Nres Freq shift r}) and (\ref{Res Freq shift r}). This can be traced back to the origin of these quantities which arise only from the electric dipole interaction of the interaction Hamiltonian given in Eq.~(\ref{Hamiltonian AF}). 

In spite of that, all velocity-dependent quantities are nonzero because now due to our effective magnetic dipole (\ref{eff magnetic moment}) as well as the electric dipole serve to probe the chirality of the mirror leading to nonzero results. All these quantities depend on 
\begin{eqnarray}
R_{nk}(\mathbf{v}_\parallel) &=& \mathrm{Im}\left\{ \left( \mathbf{d}_{nk}^* \cdot \mathbf{e}_z \right) \left[ \left( \mathbf{v}_\parallel \times \mathbf{d}_{nk} \right) \cdot \mathbf{e}_z \right] \right\} \;, \nonumber\\
&=& -\mathrm{Im}\left( d_{nk}^{\,z\,*} m_{nk}^{\,z} \right) \; , \label{ORS} 
\end{eqnarray}
where we identified in the last line the optical rotatory strength by means of the R\"ontgen interaction Hamiltonian (\ref{Roentgen}) being a $P$--odd quantity appearing naturally for any chiral molecule \cite{Butcher-Buhmann-Scheel, Molecular QED}. Therefore, we can interpret an effective circular dichroism of the transition $n\leftarrow k$ between the electric dipole and the effective magnetic moment of Eq.~(\ref{eff magnetic moment}). This means that we are observing motion--induced chirality. 

Because the optical rotatory strength $R_{nk}(\mathbf{v}_\parallel)$ (\ref{ORS}) is a pure imaginary number, it needs circularly or elliptical polarized dipoles, which contrast strongly with the required ones for its nonreciprocal counterpart $T_{nk}(\mathbf{v}_\parallel)$ (\ref{nonreciprocal ORS}). Due to its vectorial structure the noncoplanarity condition over $\mathbf{v}_\parallel$, $\mathbf{d}_{nk}$ and $\mathbf{e}_z$ and the condition $\mathbf{v}_\parallel\cdot\mathbf{d}_{nk}=0$ discussed previously are valid. The form of $R_{nk}(\mathbf{v}_\parallel)$ (\ref{ORS}) is in agreement with the Casimir--Polder potential perspective studied on Ref.~\cite{CPT-article}, where for chiral media it is required that  $\mathbf{d}_{nk}\otimes\mathbf{m}_{kn}$ must be a pure imaginary number as part of the crossed polarisability. In Table \ref{Tab-Symmetries} we resume the behavior that a perfectly reflecting chiral mirror and its dipole moments $\mathbf{d}_{nk}$ must have under $P$ and $T$ transformations.
 
Similarly to our discussion for the previous nonreciprocal materials, we also observe that these quantities could increase or decrease the corresponding contribution depending on which handedness for the mirror is used. This is a significant difference with the nonreciprocal mirror of Sec. \ref{Nonreciprocal mirror}, whose quantities lack this freedom. The final comment for this Section is that the frequency shift of Eq.~(\ref{Shift r v nret prcm}) reproduces  the $z^{-3}$ dependence of Ref.~\cite{Crosse-Fuchs-Buhmann} obtained through perturbation theory for the circularly polarized dipole $\mathbf{d}_{nk}=d_{nk}/\sqrt{2}(0,1,\mathrm{i})^\top$.


\subsection{Isotropic Chiral medium}\label{Chiral medium}


For these materials, the electric and magnetic field are connected with the displacement field $\hat{\mathbf{D}}\left(\mathbf{r};\omega\right)$ and the magnetic field $\hat{\mathbf{H}}\left(\mathbf{r};\omega\right)$ via the Boys-Post constitutive relationships 
\cite{Butcher-Buhmann-Scheel,Buhmann-Butcher-Scheel,Lakhtakia,Lindell1}
\begin{eqnarray}
\hat{\mathbf{D}}\left(\mathbf{r};\omega\right) &=& \varepsilon_0\varepsilon\left(\mathbf{r};\omega\right) \hat{\mathbf{E}}\left(\mathbf{r};\omega\right) -\frac{ \mathrm{i} \kappa }{ c } \hat{\mathbf{H}}\left(\mathbf{r};\omega\right) \nonumber\\
&& + \hat{\mathbf{P}}_N\left(\mathbf{r};\omega\right) -\frac{ \mathrm{i} \kappa }{ c } \hat{\mathbf{M}}_N \;, \label{Constitutive 3}
\end{eqnarray}
\begin{eqnarray}
\hat{\mathbf{B}}\left(\mathbf{r};\omega\right) &=& \mu_0\mu\left(\mathbf{r};\omega\right) \hat{\mathbf{H}}\left(\mathbf{r};\omega\right)  +  \frac{ \mathrm{i} \kappa }{ c }  \hat{\mathbf{E}}\left(\mathbf{r};\omega\right) \nonumber\\
&& + \mu_0\mu\left(\mathbf{r};\omega\right)\hat{\mathbf{M}}_N\left(\mathbf{r};\omega\right) \;, \label{Constitutive 4}
\end{eqnarray}
where $\kappa$ is the chiral cross-susceptibility, a magnetoelectric susceptibility being in general a tensor. The chiral cross-susceptibility of the medium has the effect of ``rotating" a magnetic effect to contribute towards an electric response and vice versa. The similarities and differences between the above constitutive relations and those of the 3D TI, as given by Eqs.~(\ref{Constitutive 1}) and (\ref{Constitutive 2}), can be examined from the perspectives of true and false chirality \cite{Barron 1, Barron 2,OJF-SYB2}: In this context, the above chiral medium is truly chiral, whereas the 3D TI is falsely chiral \cite{Barron 3}.

The modified Fresnel reflective coefficients for nonreciprocal media $r_{ \sigma,\sigma' }$ involved in Eq.~(\ref{GF dyadics}) can be found in Ref.~\cite{Ali} in its general form. Nevertheless, we will particularize them for the case of an interface constituted by vacuum in the upper half-space of Fig.~\ref{SETUP} and an isotropic chiral medium in the lower half-space. By employing the same notation of Sec. \ref{3D TI} 
where subscript 1 denotes the corresponding quantities in the vacuum and conversely the subscript 2 for the chiral medium, we have
\begin{eqnarray} 
r_{ \mathrm{s,s} } &=& \frac{ 1 }{ \mathcal{D} } \left[ (1+a_\mathrm{R})(1-b_\mathrm{L}) + (1+a_\mathrm{L})(1-b_\mathrm{R})\right] , \label{Fresnel 5} \\
r_{ \mathrm{s,p} } &=& \frac{ -2\mathrm{i} k_1 }{ \mathcal{D} k_1^\perp } \left( \frac{ k_2^{ \perp, \mathrm{L} } }{ k_2^{ \mathrm{L} } } - \frac{ k_2^{ \perp, \mathrm{R} } }{ k_2^{ \mathrm{R} } } \right), \label{Fresnel 6} \\
r_{ \mathrm{p,p} } &=& \frac{ 1 }{ \mathcal{D} }\left[ (1-a_\mathrm{R})(1+b_\mathrm{L}) + (1-a_\mathrm{L})(1+b_\mathrm{R})\right], \;\;\;\;\; \label{Fresnel 7} \\
r_{ \mathrm{p,s} } &=& -r_{ \mathrm{s,p} } \;, \label{Fresnel 8} 
\end{eqnarray}
where $\mathcal{D}=(1+a_\mathrm{R})(1+b_\mathrm{L}) + (1+a_\mathrm{L})(1+b_\mathrm{R})$ and
\begin{equation}
a_\mathrm{P} = \frac{ \sqrt{\varepsilon_2} k_1 k_2^{ \perp, \mathrm{P} } }{ \sqrt{\mu_2} k_1^\perp k_2^{ \mathrm{P} } } \;\; , \;\; b_\mathrm{P} = \frac{ \sqrt{\mu_2} k_1 k_2^{ \perp, \mathrm{P} } }{ \sqrt{\varepsilon_2} k_1^\perp k_2^{ \mathrm{P} } } \;,
\end{equation}
with $\mathrm{P}=\mathrm{R, \mathrm{L}}$ referring to the right and left circular polarization of the wave, respectively. Within the chiral medium the wave vectors are
\begin{eqnarray}
\left( k_2^{ \mathrm{R} }\right)^2 &=& \frac{ \omega^2 }{ c^2 } \left( \sqrt{ \varepsilon_2\mu_2 } + \kappa_2 \right)^2 \;, \label{k_R} \\
\left( k_2^{ \mathrm{L} }\right)^2 &=& \frac{ \omega^2 }{ c^2 } \left( \sqrt{ \varepsilon_2\mu_2 } -\kappa_2 \right)^2 \;, \label{k_L} \\
\left( k_2^{ \perp, \mathrm{P} } \right)^2 &=& \left( k_2^{ \mathrm{P} } \right)^2 - \left( k^\parallel_2 \right)^2 \;.
\end{eqnarray}
For the vacuum, which is at the achiral region 1 the wave vectors are the standard free-space ones
\begin{equation}
k_1=\frac{ \omega }{ c } \;\; , \;\; \left(k_1^\perp\right)^2 = k_1^2 - \left( k^\parallel_1 \right)^2 \;,
\end{equation}
where $k^\parallel = k_x^2 + k_y^2$. 

The contributions to the frequency shift (\ref{whole shift}) for the dipole (\ref{Dipole}) in the retarded limit ($\tilde{\omega}_{nk}z_A/c\gg1$) read as follows
\begin{eqnarray}
&\delta\omega^{ \mathrm{res, ret} }_{nk} (\mathbf{r}_A)& = \frac{ \mu_0 \tilde{\omega}_{nk}^2 |\mathbf{d}_{nk}^{\parallel}|^2 }{ 16 \pi \hbar z_A } \nonumber\\
&& \times\left[\cos\left( \frac{ 2\tilde{\omega}_{nk}z_A }{ c } \right)\mathrm{Re}\left(r^{ \mathrm{ret} }_{ \mathrm{ p,p} } - r^{ \mathrm{ret} }_{ \mathrm{ s,s} }\right)\right. \nonumber\\ 
&& \left. +\sin\left( \frac{ 2\tilde{\omega}_{nk}z_A }{ c } \right) \mathrm{Im}\left(r^{ \mathrm{ret} }_{ \mathrm{ p,p} } - r^{ \mathrm{ret} }_{ \mathrm{ s,s} }\right) \right] \nonumber\\
&& -\frac{ \mu_0 \tilde{\omega}_{nk} |d_{nk}^{\,z}|^2 }{ 8 \pi \hbar c z_A^2 } \left[ \mathrm{Re}(r^{ \mathrm{ret} }_{ \mathrm{ p,p} })\sin\left( \frac{ 2\tilde{\omega}_{nk}z_A }{ c } \right)  \right. \nonumber\\
&&  \left. +\mathrm{Im}(r^{ \mathrm{ret} }_{ \mathrm{ p,p} }) \cos\left( \frac{ 2\tilde{\omega}_{nk}z_A }{ c } \right) \right] 
\;, \label{Shift r res ret chiral} 
\end{eqnarray}
\begin{eqnarray}
&&\delta\omega^{ \mathrm{nres, ret} }_{nk} (\mathbf{r}_A) = \frac{c^2}{32\pi^2 \hbar \varepsilon_0 \tilde{\omega}_{nk}^2 z^5_A} \nonumber\\
&& \times \left[\frac{|\mathbf{d}_{nk}^{\parallel}|^2}{4} \mathrm{Im}\left(r^{ \mathrm{ret} }_{ \mathrm{ s,s} } - r^{ \mathrm{ret} }_{ \mathrm{ p,p} }\right) -|d_{nk}^{\,z}|^2\mathrm{Im}(r^{ \mathrm{ret} }_{ \mathrm{ p,p} })\right] \nonumber\\
&& -\frac{ c }{ 32 \pi^2 \hbar \varepsilon_0 \tilde{\omega}_{nk} z_A^4 } 
%
\left[ \frac{|\mathbf{d}_{nk}^{\parallel}|^2}{2} \mathrm{Re}\left(r^{ \mathrm{ret} }_{ \mathrm{ s,s} } - r^{ \mathrm{ret} }_{ \mathrm{ p,p} }\right) \right. \nonumber\\
&& \left. -|d_{nk}^{\,z}|^2\mathrm{Re}(r^{ \mathrm{ret} }_{ \mathrm{ p,p} })\right] \;, \label{Shift r nres ret chiral}
\end{eqnarray}
\begin{eqnarray}
&\delta\omega^{ \mathrm{ ret} }_{nk} (\mathbf{r}_A, \mathbf{v})& = \frac{ \tilde{\omega}_{nk}^2 R_{nk}(\mathbf{v}_\parallel) }{ 4 \pi \hbar \varepsilon_0 c^2 z_A } \left[ \mathrm{Re}(r^{ \mathrm{ret} }_{ \mathrm{ s,p} })\sin\left( \frac{ 2\tilde{\omega}_{nk}z_A }{ c } \right) \right. \nonumber\\
&& \left. +\mathrm{Im}(r^{ \mathrm{ret} }_{ \mathrm{ s,p} })\cos\left( \frac{ 2\tilde{\omega}_{nk}z_A }{ c } \right) \right]  
%
%
\;, \label{Shift r v ret chiral}
\end{eqnarray}
where $S_{nk}(\mathbf{v}_\parallel)=0$ from Eq.~(\ref{OLS}) was used and the optical rotatory strength $R_{nk}(\mathbf{v}_\parallel)$ was defined in Eq.~(\ref{ORS}), and both of them will be subsequently employed. In this limit, we found that for $\kappa_2\in\mathbb{C}$, $r_{ \mathrm{ s,p} }^{ \mathrm{ret} }$ will not vanish. For this reason, we consider the more general case of $\varepsilon_2,\kappa_2\in\mathbb{C}$ to analyze the chiral effects  in this retarded limit. The retarded limit of the Fresnel coefficients are given in Appendix \ref{B}. Again, the dominant contribution to the frequency shift (\ref{Shift r v ret chiral}) comes from the electric interaction $\mathbf{d}_{mn} \cdot \mathbf{ \hat{E} }(\mathbf{r}_A)$ as in the perfectly reflecting nonreciprocal mirror (\ref{Shift r v ret prnm}), the 3D TI (\ref{Shift r v ret TI}) and the perfectly reflecting chiral mirror (\ref{Shift r v ret prcm}). For the nonretarded limit ($\tilde{\omega}_{nk}z_A/c\ll1$), we obtain
\begin{eqnarray}
&\delta\omega^{ \mathrm{res, nret} }_{nk} (\mathbf{r}_A)& = \frac{ \mu_0 \tilde{\omega}_{nk}^2  }{ 8 \pi \hbar }\left[|\mathbf{d}_{nk}^{\parallel}|^2  \left( \frac{ r^{ \mathrm{nret} }_{ \mathrm{s,s} } }{ 2z_A } + \frac{ r^{ \mathrm{nret} }_{ \mathrm{p,p} } c^2 }{ \tilde{\omega}_{nk}^2 z_A^3 } \right) \right. \nonumber\\
&& \left. + \frac{ c^2 r^{ \mathrm{nret} }_{ \mathrm{ p,p} }  |d_{nk}^{\,z}|^2 }{ 2 \tilde{\omega}_{nk}^2 z_A^3 } \right] \;, \label{Shift r res nret chiral} \\
&\delta\omega^{ \mathrm{nres, nret} }_{nk} (\mathbf{r}_A)& = \frac{ r^{ \mathrm{nret} }_{ \mathrm{p,p} } }{ 16 \pi \hbar \varepsilon_0 z_A^3 }\left( \frac{ |\mathbf{d}_{nk}^{\parallel}|^2 }{ 4 } + \frac{ |d_{nk}^{\,z}|^2 }{ 2 }\right) ,\qquad \label{Shift r nres nret chiral} \\
&\delta\omega^{ \mathrm{ nret} }_{nk} (\mathbf{r}_A, \mathbf{v})& = \frac{ R_{kn} \mathrm{Im}(r^{ \mathrm{nret} }_{ \mathrm{ s,p} }) }{ 8 \pi \hbar \varepsilon_0 c z_A^3 } \;, \label{Shift r v nret chiral}
%
%
%
\end{eqnarray}
where the nonretarded limit of the Fresnel coefficients are given in Appendix \ref{B}. 

For the isotropic chiral medium we appreciate that the resulting frequency shifts 
are a superposition of the corresponding quantities of the perfectly conducting mirror of Sec. \ref{Conducting mirror} and the perfectly reflecting chiral mirror of Sec. \ref{Chiral mirror} determined by the Fresnel coefficients $r_{ \sigma,\sigma' }$. As a result of that, all the previous discussions and mainly the conditions over $\mathbf{v}_\parallel$, $\mathbf{d}_{nk}$ and $\mathbf{e}_z$ discussed previously are valid here too. In Table \ref{Tab-Symmetries} we resume the behavior that an isotropic chiral medium and its dipole moments $\mathbf{d}_{nk}$ must have under $P$ and $T$ transformations, where both atom and medium exhibit true chirality. Moreover, the reflective coefficients are functions of the permittivity $\varepsilon$, the permeability $\mu$ and the chirality $\kappa$ which could help to modulate the behavior of the spectroscopical quantities. In the same spirit as the role of the velocity's sign discussed for all the previous examples, the handedness of the crossed Fresnel coefficients could increase or decrease the corresponding contribution just as found for the chiral mirror but now they are more complicated functions. 
Furthermore, in the nonretarded limit the contribution of the imaginary part of the crossed Fresnel coefficients in Eq.~(\ref{Shift r v nret chiral}) 
becomes relevant because they could be more significant than the perfectly conductor behavior. Again, due to the internal connection between the frequency shift and the Casimir--Polder force, one can also switch from an attractive to a repulsive force between atom and medium.


\section{Discussion}\label{DISCUSSION}

As a first step towards observing motion--induced chirality and quantum friction with Rydberg atoms, we will discuss the magnitude of some key observables introduced above.

The central indicator of motion--induced chirality of the associated rotatory strength~(\ref{ORS}). Table~\ref{tab:rotstrength} shows its magnitude for hydrogen in Rydberg states with a range of principal quantum numbers $\bar{n}$ for the $\ket{\bar{n}00}$ to $\ket{\bar{n}11}$ transition and a velocity parallel to the plate of $v_{\parallel}=300\,\mathrm{m}/\mathrm{s}$. 

It is interesting to compare this motion--induced rotatory strength with typical rotatory strengths of chiral molecules: For the S$_0$-S$_1$ transition in the chiral carbonyl compounds fenchone and camphor, for instance, the rotatory strength corresponds to about \SI{9.5e-55}{\coulomb^2\metre^3\per\second} and \SI{1.5e-54}{\coulomb^2\metre^3\per\second}, respectively \cite{kerber:2024,pulm:1997}.
Helicenes and their derivatives are known for comparatively large rotatory strengths, which range, depending on the specific transition selected, between around \SI{1e-56}{\coulomb^2\metre^3\per\second} and \SI{1e-53}{\coulomb^2\metre^3\per\second} for [4]-helicene, [5]-helicene and [6]-helicene \cite{brown:1971-4hel,brown:1971-5hel,brickell:1971-67hel}. The accompanying large isotropic optical rotatory dispersion of helicene derivatives qualifies them for instance as candidates for sensitive matterwave interference studies on chiral molecules as proposed in Ref.~\cite{stickler:2021}.
For the Rydberg atoms, we find that the associated rotatory strength is very high compared with regular rotatory strengths in chiral molecules, as it scales with the fourth-power of the principal quantum number.
However, this scaling is rooted in the typical quadratic scaling of electrical dipoles in Rydberg systems, and thus, in relation to the electrical properties, the associated rotatory strength is still suppressed by a factor of $v/c$ of the order of $10^{-6}$.

To estimate the magnitude of motion-induced frequency shifts as evidence for quantum friction, we consider the bare and motion-corrected frequency shifts, Eqs.~(\ref{Shift r nres nret prnm}) and (\ref{Shift r v nret prnm}), respectively, of an atom at non-retarded distances $z_A=1\,\upmu\mathrm{m}$ from the surface of a perfectly reflecting nonreciprocal mirror. 
The results for a velocity $v=300\,\mathrm{m}/\mathrm{s}$ are also displayed in Table~\ref{tab:rotstrength}. 
Note that in  \eqref{Shift r v nret prnm} we need to choose the quantization axis in the $\mathbf{v}_\parallel$-direction parallel to the plate, which results in vanishing \eqref{Shift r nres nret prnm}.
In Table~\ref{tab:rotstrength} we evaluate \eqref{Shift r nres nret prnm} therefore with respect to a quantization axis normal to the plate to allow for a useful comparison of the different frequency shifts.

Additionally, Table~\ref{tab:rotstrength} also gives values for frequency shifts in the analogue case of a chiral mirror \eqref{Shift r v nret prcm}, with adjusted transition states and quantization axis. A concrete scenario for the measurement of such frequency shifts would consist in directing a tightly focused atomic beam through a wedge-shaped gap, as done in pioneering measurements of the Casimir--Polder forces \cite{Sukenik}. In order to implement chiral mirrors, the gap walls need to either consist of nanofabricated chiral metasurfaces \cite{Voronin} or by coating with chiral molecules \cite{Suzuki2}. Recent progress in the measurement of the transition frequency of Rydberg states has shown an accuracy of a few MHz using electromagnetically induced transparency \cite{Silpa} thus static dipole related energy shifts are resolvable starting from $\bar{n}\ge60$.  Motion-corrected frequency shifts for a nonreciprocal mirror and a chiral mirror should be observable starting from $\bar{n}\ge40$ as kHz accuracy with Rydberg microwave Ramsey spectroscopy can be achieved when resorting to circular states \cite{Zou}. Frequency shifts are advantageous for measurement because frequencies can be measured with very high accuracy \cite{Keller_2016}. 
In principle, laser excitation of Rydberg states is frequency selective to the width of the corresponding transition, where sub-kHz resolution in Rydberg states can be achieved for sufficiently high $\bar{n}$. In the current application, state selective laser excitation might not be possible because of the proximity of the chiral mirror. Frequency measurement during free evolution of an atom excited to a Rydberg can be done via Ramsey interferometry. Interaction induced energy shifts lead to accumulated phase shifts, which can be projected to populations in a coherent excitation process, using other Rydberg states or a highly stable ground state as a reference. Measurements of interaction induced frequency shifts in Rydberg atoms have been implemented~\cite{Arias_2019} and proposed to be used in the application of chiral discrimination~\cite{Buhmann-Singer}.

	\begin{table*}[ht]
	\centering
		 \begin{tabular}{||c   c   c c c c ||} 
		 \hline
		  & Parameter\quad $\bar{n}\rightarrow$ & $20$ & $40$ & $60$ & $80$ \\
		 \hline
		 $R_{nk}$ && $1.291 \times 10^{-51}$ & $2.069 \times 10^{-50}$ & $1.048 \times 10^{-49}$ & $3.312 \times 10^{-49}$  \\ 
		 $ \delta\omega_{nk}$ &  & $2.918 \times 10^{7}$ & $4.677 \times 10^{8}$ & $2.369 \times 10^{9}$ & $7.487 \times 10^{9}$  \\
		 $	\delta\omega_{nk} (v)$ & (nonreciprocal)  & $9.173\times 10^{1}$ & $1.470 \times 10^{3}$ & $7.446 \times 10^{3}$ & $2.354 \times 10^{4}$  \\
		 $	\delta\omega_{nk} (v)$ & (chiral) & $1.835\times 10^{2}$ & $2.941 \times 10^{3}$ & $1.489 \times 10^{4}$ & $4.707 \times 10^{4}$  \\
		 \hline
		 \end{tabular}
		 \caption{%
         Associated rotatory strength $R_{nk}$ \eqref{ORS}, bare and motion--corrected frequency shifts \eqref{eqn:frequshiftbarenonres} and \eqref{Shift r v nret prnm} for a nonreciprocal mirror, as well as motion--induced frequency shift \eqref{Shift r v nret prcm} for a chiral mirror. 
         The nonreciprocal frequency shift is given for an electrical dipole transition via Rydberg type states $\mathbf{d}_{nk} = \braket{\bar{n}_n l_n m_n | \hat{\mathbf{d}} | \bar{n}_k l_k m_k} = \braket{\bar{n} 1 1 | \hat{\mathbf{d}} | \bar{n} 0 0}$, where the bare value is quantized in the $z$-direction, while for the motion--induced shift the quantization axis is chosen parallel to $\mathbf{v}_{\parallel}$.
         The chiral frequency shift is calculated for an electrical dipole transition via Rydberg type states $\mathbf{d}_{nk} = \braket{\bar{n}_n l_n m_n | \hat{\mathbf{d}} | \bar{n}_k l_k m_k} = \braket{\bar{n} 1 0 | \hat{\mathbf{d}} | \bar{n} 0 0}$, with quantization axis along $\mathbf{e}_y + \mathbf{e}_z$ for $\mathbf{v}_{\parallel}= v_{\parallel} \mathbf{e}_x$. 
         All values are in standard SI-units, $[R]=\textrm{C}^2 \textrm{m}^3/\textrm{s}$, $[\delta\omega]=\textrm{rad}/\textrm{s}$.
         Note that the velocity-dependent values are linear in $v/c$, with $v=|\mathbf{v}_{\parallel}|$ and the frequency shifts scale with $z_A^{-3}$; the given values are calculated for the velocity $|\mathbf{v}_{\parallel}|=300 \textrm{m}/\textrm{s}$ and at a distance $z_A=10^{-6} \textrm{m}$.} \label{tab:rotstrength}
	\end{table*}

\section{Conclusions} \label{CONCLUSIONS}

In this work macroscopic QED was used to derive expressions for the Casimir--Polder frequency shift and spontaneous decay rate for a moving atom parallel to nonreciprocal (falsely chiral) and (truly) chiral media. The former violate Lorentz's reciprocity principle meaning that time-reversal-symmetry is broken and the latter are reciprocal media breaking parity. Nonreciprocal media are included by using generalised real and imaginary parts of the Green's tensor which reduce to the standard definitions in the reciprocal case, including (truly) chiral media.

The basis of our study lies on the interaction Hamiltonian between the atom, the field and the nonreciprocal or chiral medium, which enables us to obtain an expression for the electric and magnetic field. To this end, noise currents were quantized directly yielding one set of field operators for the combined electric and magnetic field. The result for both fields allowed us to analyze the internal atomic dynamics. Through the redefinitions of the real and imaginary parts of a tensor, we provide general expressions for the atomic decay rate and the frequency shift, which can be split into two kinds of terms. One depends only on the position, and another one depends on the position and the velocity of the atom. Also we found that such terms can be split into resonant and nonresonant contributions. These results reproduce and extend the previous results for a static atom of Ref.~\cite{Fuchs-Crosse-Buhmann} to a polarizable atom moving parallel to nonreciprocal or chiral media.

By regarding only a moving non-magnetic atom we analyzed five different materials: a perfectly conducting mirror, a perfectly reflecting nonreciprocal mirror, a strong three-dimensional topological insulator, a perfectly reflecting chiral mirror and an isotropic chiral medium. We find different power laws for all these materials in the nonretarded limit and oscillating functions whose amplitudes decay with the distance in the retarded one.

Let us begin by summarising the results for the static 
frequency shift of nonreciprocal media. By employing an arbitrary dipole moment, we are able to generalize the results of Ref.~\cite{Fuchs-Crosse-Buhmann}. Finding that for the perfectly conducting mirror there is no difference between any special polarizations because its frequency shift depends only on the magnitude of the dipole moments, i.e. all of them will always contribute. More interesting are the results for the nonreciprocal mirror, because its 
frequency shift depends on $\mathrm{Im}(d_{nk}^{\,x}d_{nk}^{\,y\,*})$ which rewritten as Eq.~(\ref{d x d* ez}) can be identified as the $z$ component of the geometric magnetic field $\mathcal{B}_z$ associated with photoionization of chiral molecules \cite{Ayuso-Ordonez-Smirnova,Ordonez-Smirnova}. This constitutes the first bridge between the physics of chirality and nonreciprocity obtained in this work: nonreciprocal media couple to the falsely chiral response of the molecule. For the strong 3D TI we found that its 
frequency shift is a superposition of the perfectly conducting mirror and the nonreciprocal mirror ones but modulated by the Fresnel coefficients. The latter provide the possibility to modulate the behavior of the spectroscopical quantities by adjusting the permittivity $\varepsilon$ and mainly the topological parameter $\Delta$ because it can be negative for certain strong 3D TIs. All these results reproduce the same algebraic power laws and fading-out functions found in Ref.~\cite{Fuchs-Crosse-Buhmann} for the nonretarded and retarded limits respectively. 
 
Now we move on with the results for the static 
frequency shift of the chiral media. For the perfectly reflecting chiral mirror and the isotropic chiral medium these spectroscopical quantities are identically zero. According to Curie's principle \cite{Curie} these results are reasonable because our chosen dipole moments do not share the same symmetry. The square of the transition dipoles constitute an achiral response that is unable to resolve the chirality present in both materials. So, to probe the chirality in this static case one should employ a molecule with a magnetic moment as well as an electric one. 

On the other hand, we have the velocity-dependent 
frequency shift of the nonreciprocal media. After employing again an arbitrary dipole moment, we find that for the perfectly conducting mirror the relevant vectorial structure is $S_{nk}(\mathbf{v}_\parallel)=\mathrm{Im}\left[ \left( \mathbf{d}_{nk}^* \cdot \mathbf{e}_z \right) \left( \mathbf{d}_{nk} \cdot \mathbf{v}_\parallel \right) \right]$, which is zero from the Maxwell equations conditions leading to a vanishing shift and guarantees the $P$--even and $T$--even behavior of this material. 
For the nonreciprocal mirror the vectorial structure can be understood in terms of the nonreciprocal optical rotatory strength $T_{nk}(\mathbf{v}_\parallel)=\mathrm{Re}\left\{ \left( \mathbf{d}_{nk}^* \cdot \mathbf{e}_z \right) \left[ \left( \mathbf{v}_\parallel \times \mathbf{d}_{nk} \right) \cdot \mathbf{e}_z \right] \right\}$ given by Eq.~(\ref{nonreciprocal ORS}) which needs a superposition of linear polarized dipoles to not vanish. Because of its similarity with the optical rotatory strength of chiral media, this constitutes the second bridge found in this work between the physics of chirality and nonreciprocity. Then, we analyzed 
the frequency shift of a strong 3D TI. In this case, we found again that its 
frequency shift is a superposition of the perfectly conducting mirror and the nonreciprocal mirror ones but modulated by the Fresnel coefficients. However, the sign of the velocity could lead to an enhancement or decrease in the effect for the three materials, but for the strong 3D TI the topological parameter $\Delta$ is another quantity that has this feature. Algebraic power laws in the atom's distance $z_A$ to the interface are provided for the three materials in the nonretarded limits, which could be useful for experimental verification. All these results are consistent with Curie's principle because $T_{nk}$ is a $T$--odd quantity 
and also the strong 3D TI breaks time-reversal symmetry via the topological parameter $\Delta$ being a $T$--odd quantity. For this reason, this study gains significance for ongoing investigations into dark matter detection, where axions emerge as a promising candidate \cite{Sikivie}. In this pursuit, condensed matter physics offers novel avenues through TIs \cite{Nenno et al}, antiferromagnetically doped TIs \cite{Marsh et al, Jan Schuette}, or multiferroics \cite{Roising et al}, utilizing the magnetoelectric effect which underpins our results.

Regarding the velocity-dependent 
frequency shift of the chiral media. Let us first describe the results for the perfectly reflecting chiral mirror. Here we found that this spectroscopical quantity is determined by $R_{nk}(\mathbf{v}_\parallel)=\mathrm{Im}\left\{ \left( \mathbf{d}_{nk}^* \cdot \mathbf{e}_z \right) \left[ \left( \mathbf{v}_\parallel \times \mathbf{d}_{nk} \right) \cdot \mathbf{e}_z \right] \right\}$ which is the the optical rotatory strength \cite{Butcher-Buhmann-Scheel, Molecular QED} given by Eq.~(\ref{ORS}). Therefore, we can interpret an effective circular dichroism of the transition $n\leftarrow k$ between the electric dipole $\mathbf{d}_{nk}$ and the effective magnetic moment $\mathbf{m}_{nk} = - \mathbf{v}_\parallel \times \mathbf{d}_{nk}$. Although we did not employ a chiral source, the rectilinear uniform motion of our electric dipole induces an effective magnetic moment through its velocity as found in Eq.~(\ref{eff magnetic moment}), therefore a chiral source is indirectly used to probe the chirality of this mirror explaining the non-zero results, i.e., here occurs a motion--induced (true) chirality. Lastly, for the isotropic chiral medium its 
frequency shift is a superposition of the perfectly conducting mirror and the chiral mirror ones but modulated by the Fresnel coefficients. For both chiral materials the sign of the velocity and the handedness of the media could lead to an enhancement or decrease in the effect.  Algebraic power laws in the atom's distance $z_A$ to the interface are provided for these two materials in the nonretarded limits, which could be useful for experimental verification. Nevertheless, for the isotropic chiral medium, the complex character of the Fresnel coefficient can enhance the power law of the distance to the surface. All these results are consistent with Curie's principle because $R_{nk}$ is a $P$--odd quantity, if circular or elliptical polarizations are selected, and also the isotropic chiral medium breaks the parity symmetry via its chirality $\kappa$ being a $P$--odd quantity.  

It is worth mentioning a few comments before closing this section. First, for the current interface $\Delta$ is proportional to the fine structure constant $\alpha$ as mentioned in Eq.~(\ref{DELTA}). Nevertheless, we could replace the strong 3D TI by a magnetoelectric such as TbPO$_4$ whose $\Delta=0.22$ \cite{TbPO4,Rivera} increases the influence of the non-reciprocal medium. However we focus on the current work only on strong 3D TIs. Second, the present work provides general results for 
the frequency shifts either resonant or nonresonant but we remark that all the concrete results for the five materials are valid only for constant permittivities and non-magnetic strong 3D TIs, i.e. for the five materials  their susceptibilities do not have frequency dependence. Third, due to a more exhaustive analysis of the frequency derivative involved in the velocity-dependent decay rates, we have left their study for these materials for future work. Fourth, of course, the study of the perpendicular motion ($v_z\neq0$) can be investigated, but it has problems and we will leave it for further studies, which resembles the classical situation of transition radiation studied previously in Ref.~\cite{OJF-SYB}. Finally, all these results are obtained within the Markovian framework, meaning that they are only valid for times that verify the constraints described in Ref.~\cite{Klatt-Kropf-Buhmann}.

\section{Acknowledgments}
We would like to thank Stephen Barnett, Lawrence D. Barron, Robert Bennett, Quentin Bouton, Carsten Henkel, Francesco Intravaia, Athanasios Laliotis, Daniel Reiche and Fumika Suzuki for discussions. We also acknowledge the advice of K. Hartwig for editing the manuscript. O. J. F. has been supported by the postdoctoral fellowships CONACYT-770691, CONACYT-800966. The authors gratefully acknowledge funding by the Deutsche Forschungsgemeinschaft Project No. 328961117 -- SFB 1319 ELCH.

\appendix

\section{Fresnel reflection coefficients for a vacuum-strong 3D TI interface} \label{A}
To the purpose of this article be self-contained in this appendix, we provide the necessary details of the reflective coefficients in the retarded and nonretarded limits for nonreciprocal media. 

The Fresnel coefficients for an interface constituted by a 3D topological insulator with $\mu_2=1$ and vacuum in the retarded limit ($\tilde{\omega}_{nk}z_A/c\gg1$) whose main contribution comes from the region due to small $k_\parallel$ are \cite{Fuchs-Crosse-Buhmann}: 
\begin{eqnarray}
r^{ \mathrm{ret} }_{ \mathrm{s,s} } &=& \frac{ 1 - \varepsilon - \Delta^2 }{ (1 + \sqrt{\varepsilon_2} )^2 + \Delta^2 } \;, \\
r^{ \mathrm{ret} }_{ \mathrm{p,s} } &=& \frac{ -2\Delta }{ (1 + \sqrt{\varepsilon_2})^2 + \Delta^2 } \;,\\
r^{ \mathrm{ret} }_{ \mathrm{p,p} } &=& - r^{ \mathrm{ret} }_{ \mathrm{s,s} } \;,\\
r^{ \mathrm{ret} }_{ \mathrm{s,p} } &=& r^{ \mathrm{ret} }_{ \mathrm{p,s} } \;,
\end{eqnarray}
where $\Delta$ is given by Eq.~(\ref{DELTA}). \\

On the other hand, these coefficients in the nonretarded limit ($\tilde{\omega}_{nk}z_A/c\ll1$) whose leading contribution comes from the region due to large $k_\parallel$ are \cite{Fuchs-Crosse-Buhmann}: 
\begin{eqnarray}
r^{ \mathrm{nret} }_{ \mathrm{s,s} } &=& \frac{ -\Delta^2 }{ 2(\varepsilon_2 +1) + \Delta^2 } \;, \\
r^{ \mathrm{nret} }_{ \mathrm{s,p} } &=& \frac{ -2\Delta }{ 2(\varepsilon_2 +1) + \Delta^2 } \;, \\
r^{ \mathrm{nret} }_{ \mathrm{ p,p} } &=& \frac{ 2(\varepsilon_2 -1) + \Delta^2 }{ 2(\varepsilon_2 +1) + \Delta^2 } \;, \\
r^{ \mathrm{nret} }_{ \mathrm{s,p} } &=& r^{ \mathrm{nret} }_{ \mathrm{p,s} } \;.
\end{eqnarray}

\section{Fresnel reflection coefficients for the vacuum-isotropic chiral medium interface} \label{B}

The corresponding Fresnel coefficients for an interface constituted by an isotropic chiral medium and vacuum in the retarded limit ($\tilde{\omega}_{nk}z_A/c\gg1$) whose main contribution comes from the region due to small $k_\parallel$ are \cite{Rapp}: 
\begin{equation}
\begin{aligned}
r^{ \mathrm{ret} }_{ \mathrm{s,p} } &= \frac{ 2 \mathrm{i}\, ( k_{\mathrm{L}}|k_{\mathrm{R}}| - k_{\mathrm{R}}|k_{\mathrm{L}}| ) }{ 2\left(\frac{ \varepsilon + \mu }{n_r}\right)(k_{\mathrm{L}}k_{\mathrm{R}} + |k_{\mathrm{R}}k_{\mathrm{L}}|) + ( k_{\mathrm{L}}|k_{\mathrm{R}}| + k_{\mathrm{R}}|k_{\mathrm{L}}| ) } ,\\\\\
r^{ \mathrm{ret} }_{ \mathrm{p,s} } &= - r^{ \mathrm{ret} }_{ \mathrm{s,p} } \,, \\
r^{ \mathrm{ret} }_{ \mathrm{s,s} } &= \frac{ 2( k_{\mathrm{L}}k_{\mathrm{R}} - |k_{\mathrm{R}}k_{\mathrm{L}}| ) - \left(\frac{ \varepsilon - \mu }{n_r}\right)(k_{\mathrm{L}}|k_{\mathrm{R}}| + k_{\mathrm{R}}|k_{\mathrm{L}}|) }{ 2(k_{\mathrm{L}}k_{\mathrm{R}} + |k_{\mathrm{R}}k_{\mathrm{L}}|) + \left(\frac{ \varepsilon + \mu }{n_r}\right)( k_{\mathrm{L}}|k_{\mathrm{R}}| + k_{\mathrm{R}}|k_{\mathrm{L}}| ) } , \\\\\
r^{ \mathrm{ret} }_{ \mathrm{p,p} } &= \frac{ 2( k_{\mathrm{L}}k_{\mathrm{R}} - |k_{\mathrm{R}}k_{\mathrm{L}}| ) + \left(\frac{ \varepsilon - \mu }{n_r}\right)(k_{\mathrm{L}}|k_{\mathrm{R}}| + k_{\mathrm{R}}|k_{\mathrm{L}}|) }{ 2(k_{\mathrm{L}}k_{\mathrm{R}} + |k_{\mathrm{R}}k_{\mathrm{L}}|) + \left(\frac{ \varepsilon + \mu }{n_r}\right)( k_{\mathrm{L}}|k_{\mathrm{R}}| + k_{\mathrm{R}}|k_{\mathrm{L}}| ) } ,\\\
\end{aligned}
\end{equation}
where we recall the definitions for $k_\mathrm{R}$ and $k_\mathrm{L}$ given in Eqs.~(\ref{k_R}) and (\ref{k_L}). To obtain these coefficients, the approximations $k^{\|}=0$ and 
\begin{equation}\label{ret approx hs}
k^\perp\simeq n_r \omega/c =\sqrt{\varepsilon(\omega)\mu(\omega)} \omega/c
\end{equation}
were applied to their general expressions given in Ref.~\cite{Ali}.

On the other hand, these coefficients in the nonretarded limit ($\tilde{\omega}_{nk}z_A/c\ll1$) whose leading contribution comes from the region due to large $k_\parallel$ are  \cite{Butcher-Buhmann-Scheel}: 
\begin{eqnarray}
r^{ \mathrm{nret} }_{ \mathrm{s,s} } &=& \frac{ \varepsilon_2 \mu_2 - \kappa_2^2 - \varepsilon_2 + \mu_2 -1 }{ \varepsilon_2 \mu_2 - \kappa_2^2 + \varepsilon_2 + \mu_2 +1 } \;, \\
r^{ \mathrm{nret} }_{ \mathrm{p,s} } &=& \frac{ - 2 \mathrm{i} \kappa_2 }{ \varepsilon_2 \mu_2 - \kappa_2^2 + \varepsilon_2 + \mu_2 +1 }\;, \\
r^{ \mathrm{nret} }_{ \mathrm{p,p} } &=& \frac{ \varepsilon_2 \mu_2 - \kappa_2^2 + \varepsilon_2 - \mu_2 -1 }{ \varepsilon_2 \mu_2 - \kappa_2^2 + \varepsilon_2 + \mu_2 +1 } \;, \\
r^{ \mathrm{nret} }_{ \mathrm{s,p} } &=& - r^{ \mathrm{nret} }_{ \mathrm{p,s} } \;.
\end{eqnarray}


\end{document}